\documentclass{aastex}
\usepackage{times}
\usepackage{emulateapj5}

\def\MAP{\hbox{\sl MAP~}}
\def\rms{\hbox{\it rms\/}}
\def\deg{\hbox{$^\circ$}}
\def\etal{\it et al.}

\slugcomment{The Astrophysical Journal, in press}
\shorttitle{MAP Optical Design and Characterization}
\shortauthors{Page et al.}

\begin{document}

\title{The Optical Design and Characterization of the Microwave Anisotropy Probe}

\author{L. Page\altaffilmark{1}, C. Jackson\altaffilmark{2},  
C. Barnes\altaffilmark{1}, C. Bennett\altaffilmark{3},
M. Halpern \altaffilmark{4}, G. Hinshaw \altaffilmark{3}, 
N. Jarosik \altaffilmark{1}, A. Kogut \altaffilmark{3}, 
M. Limon \altaffilmark{1,3}, S. S. Meyer \altaffilmark{5},
D. N. Spergel \altaffilmark{6}, G. S. Tucker \altaffilmark{7},
D. T. Wilkinson \altaffilmark{1}, E. Wollack \altaffilmark{3},
E. L. Wright \altaffilmark{8}}

\altaffiltext{1}{Dept. of Physics, Princeton University, Princeton, NJ 08544}
\altaffiltext{2}{Code 556, Goddard Space Flight Center, 
                 Greenbelt, MD 20771}
\altaffiltext{3}{Code 685, Goddard Space Flight Center, 
                 Greenbelt, MD 20771}
\altaffiltext{4}{Dept. of Physics, Univ. Brit. Col., Vancouver, B.C., 
                 Canada V6T 1Z4}
\altaffiltext{5}{Astronomy and Physics, University of Chicago, 
                 5640 South Ellis Street, LASP 209, Chicago, IL 60637}
\altaffiltext{6}{Dept of Astrophysical Sciences, Princeton University,
                 Princeton, NJ 08544}
\altaffiltext{7}{Dept. of Physics, Brown University, Providence, RI 02912}
\altaffiltext{8}{Astronomy Dept., UCLA, Los Angeles, CA 90095}
\email{page@princeton.edu}


\begin{abstract}
The primary goal of the {\sl MAP} satellite, now in orbit, is
to make high fidelity polarization sensitive maps of the full sky in
five frequency bands between 20 and 100 GHz. From these maps we will
characterize the properties of the cosmic microwave background (CMB)
anisotropy and Galactic and extragalactic emission on angular scales
ranging from the effective beam size, $<0\fdg23$, to the full sky.
{\sl MAP} is a differential microwave radiometer. Two back-to-back shaped 
offset Gregorian telescopes feed two mirror symmetric arrays
of ten corrugated feeds. We describe the prelaunch design and characterization
of the optical system, compare the optical models to the measurements, 
and consider multiple possible sources of systematic error. 

\end{abstract}

\section{Introduction}

The Microwave Anisotropy Probe (\MAP) was designed to produce an
accurate full-sky map of the angular variations in microwave flux 
\footnote{{\sl MAP} is sensitive to only the anisotropy and is
insensitive to the ``absolute'' or isotropic flux component.},
in particular the cosmic microwave background 
(CMB) \cite{Bennett03}.
The scientific payoff from studies of the CMB anisotropy has driven
specialized designs of instruments and observing strategies since
the CMB was discovered in 1965 \cite{PW}. 
Experiments that have detected signals consistent with the CMB anisotropy 
include\footnote{References that
summarized  multiple measurements or that emphasized the 
design of an experiment were chosen.}
ground based telescopes using differential or beam synthesis techniques: 
IAB \cite{PicCal}, PYTHON \cite{coble99,Platt97}, VIPER \cite{peterson00}, 
SASK \cite{Wollack97}, SP \cite{gundersen95}, TOCO \cite{miller02}, 
IAC/Bartol \cite{romeo01}, TENERIFE \cite{hancock97},
OVRO/Ring \cite{myers93}, OVRO \cite{leitch00}; interferometers: 
CBI \cite{padin01}, CAT \cite{baker99}, DASI \cite{leitch01},
IAC \cite{harrison00}, VSA \cite{Watson02}; balloons: FIRS \cite{ganga93}, 
ARGO \cite{debernardis94},  MAX \cite{lim96}, QMAP \cite{Devlin98},
MAXIMA \cite{Lee99}, MSAM \cite{wilson00}, BAM \cite{tucker97}, 
BOOMERanG \cite{Crill02}, ARCHEOPS \cite{Benoit02}; and the {\sl COBE/DMR} satellite \cite{Smoot90}. 
Of these, only DMR has produced a full sky map. 

For \MAP, the experimental challenge was to design a mission that measures
the temperature difference between two pixels of sky separated by
180\deg  as accurately and precisely as the difference between
two pixels separated by 0\fdg25. Additionally, we required that the 
measurements be as uncorrelated with each other as possible in order 
to make detailed 
statistical analyses of the maps tractable and so that a simple list of 
pixel temperatures and statistical weights alone would accurately
describe the sky. We also required that the systematic
error on any mode in the final map, {\it before} modeling, be $<4~\mu$K 
of the target sensitivity of $20~\mu$K per 
$3.2\times 10^{-5}$~sr pixel. Equivalently, the systematic 
variance should be $<5$\% of the target noise variance. 

The components of the \MAP  mission---receivers, optics, scan strategy,
thermal design, electrical design, and attitude control---all work together. 
Without any
one of them, the mission would not achieve the goals set out above.
One guiding philosophy is that a differential measurement
with a symmetric instrument is highly desirable as discussed, for
example, by Dicke (1968). The reason is that differential outputs are, to
first order, insensitive to changes in the satellite temperature or radiative
properties. This is especially important for 
variations on time scales up to $\sim$ 1 hr, the precession period of 
{\sl MAP's} compound spin. The philosophy is
naturally suited to the need to detect the celestial signal well above
the $1/f$ knee of the HEMT amplifiers as discussed in a companion paper
\cite{Jarosik03}. Other key aspects of the design include 
simplicity, heritage of major 
components, the minimization of moving parts, and a single mode of
operation.
 
In this paper, we discuss the design of the optics, how the design 
relates to the science goals, and how the optical response 
is quantified.  To put the final
design into perspective, some of the trade-offs are discussed.
It is worth keeping in mind that our knowledge of the
optics is one of the limiting uncertainties for \MAP. 

\subsection{Design Outline}

{\sl MAP} uses a pair of back-to-back offset shaped Gregorian
telescopes that focus celestial radiation onto ten pairs of back-to-back
corrugated feeds as shown in Figure~\ref{fig:sideview} and in 
Bennett {\it et al.} (2003). The feeds are designed to 
accept radiation in five
frequency bands between 20 and 100 GHz. Table~\ref{tab:measbeams}
shows the band conventions.  Two linear orthogonal polarizations
from each feed are selected by an orthomode transducer.
Each polarization is separately amplified and detected. 

The primary design considerations were as follows:

\begin{itemize}

\item The optical system plus thermal radiators
must fit inside the 2.74~m diameter 
MIDEX fairing, and have $>25~$Hz resonant
frequency. The less massive the structure, the less support structure is
required, and the easier it is to thermally isolate the system.
The mass limit of the entire payload is 840~kg.
The center of mass of the system must survive launch accelerations 
which can attain 12g. Some components experience significantly
higher accelerations.

\item The main beam diameter must be $<0.3^{\circ}$, characterized to $-30~$dB
in flight, and computable to high accuracy. The cross polarization must 
be $<-20~$dB so that the polarization of the anisotropy in 
the CMB may be accurately determined.
The focal plane must accommodate 10 dual polarization feeds in five
frequency bands and allow for all the waveguide attachments.

\item The sidelobes must be $<-55~$dBi \footnote{The unit dBi refers to
the gain of a system relative to an isotropic emitter; dB generically
refers to just relative gain.} at
the position of the Sun, well characterized, and theoretically
understood. In addition, the response to the galaxy through the 
sidelobes must be $<1$\% of the main beam response. A premium was placed
on the computability of the sidelobes and the absence of cavities 
or enclosures in which standing waves could be set up. Thus on-axis 
designs and designs with support structure that might scatter radiation   
were not considered. 

\item The reflectors and the optical cavity (see Figure~\ref{fig:sideview}) 
must be thermally isolated 
from the spacecraft and must radiatively cool to $\sim 70~$K in flight. 
The reflector
surfaces must have $<1$\% microwave emissivity, must not build up charge 
on the surface, and, along with the feeds, must be able to withstand 
direct illumination by the Sun down the optical boresight for brief periods.

\begin{figure*}[tb]
\epsscale{1.2}
\plotone{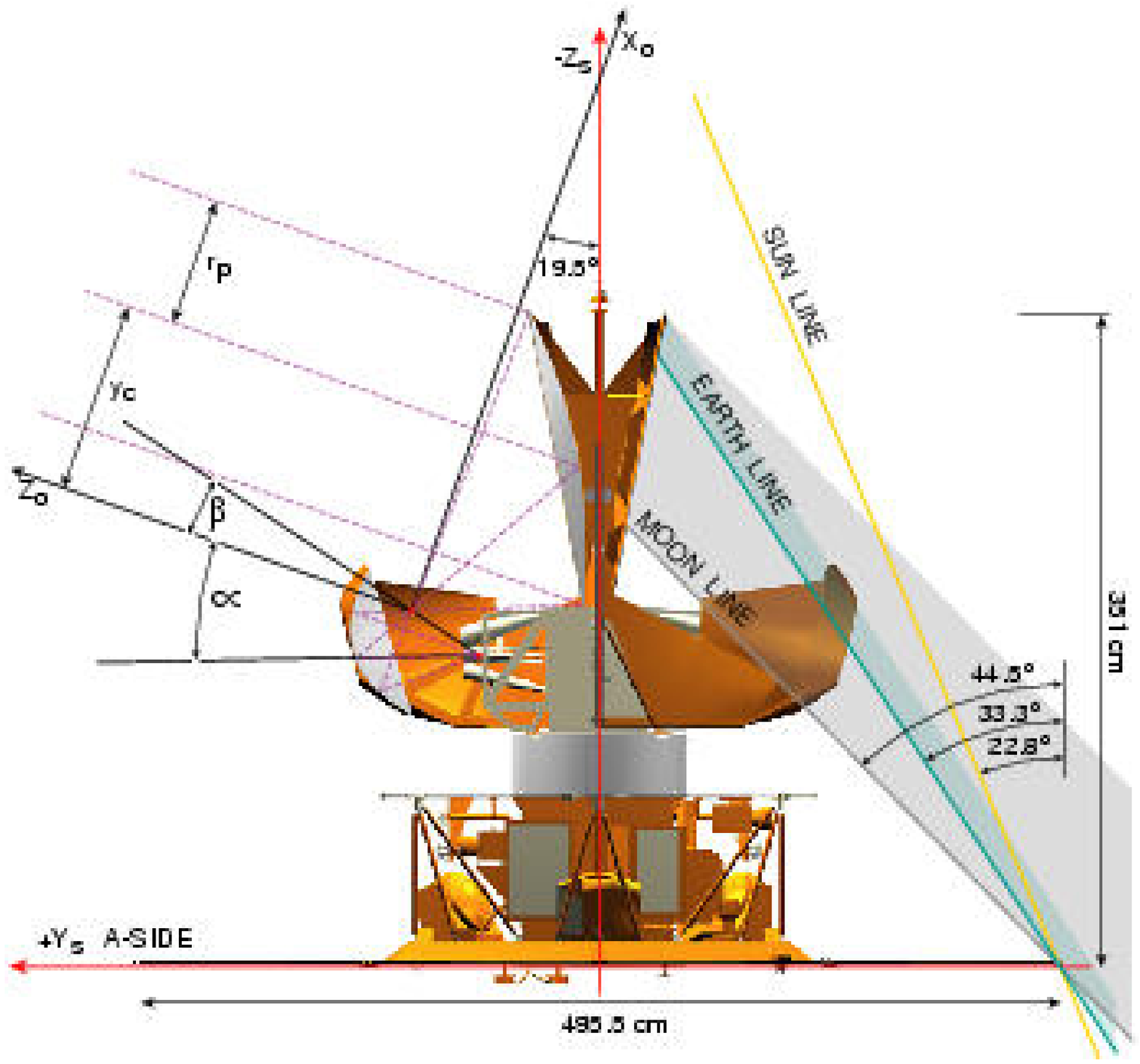}
\caption{Side view of the {\sl MAP} optical system. The left side shows a
cut-away view illustrating the relation of the microwave feeds
to the optical system. We call the region around the feeds and secondary 
the optical cavity. The dashed lines represent three rays from a
hypothetical feed at the center of the focal surface. The white conical
structures are the feeds which are held in the FPA. The centrally
located white gamma-alumina cylinder (diameter: 94~cm, height: 33~cm, 
thickness: 0.3~cm) thermally isolates the cold ($\approx70~$K) optics 
from the warm ($\approx290~$K) 
spacecraft. The parameters that describe a Gregorian are also
shown. The right side shows how the Earth, Moon and Sun illuminate {\sl MAP}
at their minimum angles of incidence. In this view, the radiators are
seen edge on. Platinum resistive thermometers are located
on the primaries at $z_s=-350$~cm (top) and $z_s=-279$~cm (middle)
and secondaries at $z_s=-214,-170,\& -138$~cm. 
Other views of the S/C may be
found in Bennett {\it et al.} (2003), Jarosik {\it et al.} (2003), and 
Barnes {\it et al.} (2002).}
\label{fig:sideview}
\end{figure*}
 
\end{itemize}

\subsection{Terminology and Conventions}

Throughout this paper the following terminology is used:
The thermal reflector system (TRS) consists of the primary and
secondary mirrors of each telescope, the structure that supports them,
and the passive thermal radiators as shown in Figure~\ref{fig:sideview}.
The structure that holds the 
feeds and the cold end of the receiver chains is called the focal plane
assembly, or FPA. The TRS fits over the FPA. 
A reflector evaluation unit (REU), which comprises half the TRS optics, was
built to assess the design, for outdoor beam mapping, and as a ground
reference unit. ``S/C'' is used for spacecraft.

In quantifying the \MAP's response, where possible, the
definitions of Kraus (1986) are used. The normalized antenna 
response to power is 
\begin{equation}
B_n({\theta,\phi}) =
{|\psi(\theta,\phi)|^2\over|\psi|^2_{max}},
\end{equation}
where $\psi$ is the scalar electric field in units of
${\rm W^{1/2}m^{-1}}$ and is evaluated at a
fixed distance from the source. The full-width-half-max of a
symmetric beam is the angle at which $B_n(\theta_{fwhm}/2) = 1/2$
(or $-3$~dB). When the beam is asymmetric, separate
$\theta_{fwhm}$ are quoted or the geometric mean of the two 
$\theta_{fwhm}$ is used. At the output of any lossless 
antenna system, one measures the power in watts given by

\begin{equation} 
W = {1\over
2}\int_{\Omega}\int_{\nu}A_e(\nu ) S_{\nu}(\theta,\phi)
B_n({\nu,\theta,\phi})d\,\Omega d\,\nu,
\label{eq:power}
\end{equation}
where $A_e$ is the effective area of the antenna and 
$S_{\nu}(\theta,\phi)$ is the brightness of the sky.  
The directivity is
\begin{eqnarray}
\lefteqn{ D_{max} \equiv {|\psi|_{max}^2\over |\psi_{avg}|^2}=
{4\pi |\psi|_{max}^2\over 
\int_{\Omega}|\psi(\theta,\phi)|^2 d\Omega}= }\hspace{0.75in} \nonumber \\
& & {4\pi\over\int B_n({\theta,\phi})d\Omega}
= {4\pi\over \Omega_{A}} = 
{4\pi A_{e}\over n\lambda^2},
\end{eqnarray}
where $\Omega_{A}$ is the total solid angle of the
normalized antenna pattern, $n$ is the number of radiative modes,
and $|\psi_{avg}|^2$ is 
the total power averaged over the sphere. For single moded systems such as
{\sl MAP}, $n=1$. When there are no losses in the telescope,
$G(\theta,\phi)=D_{max}\,B_n({\theta,\phi})$ and the 
directivity is the maximum gain.

The maximum gain, $G_m$, is sometimes just called the gain or ``the gain
above isotropic.'' It can be understood by considering the flux
(power/area) from an isotropic emitter of total power $P$. At a distance
$r$, the flux is $P/4\pi r^2$ and the gain is unity (0 dBi). If instead
the power were emitted by a feed of gain $G_m$ the flux at the maximum
would be $I = G_m P/ 4\pi r^2~{\rm W/m^2}$.
In other words, if one measures the field at a distance
$r$ from the feed then
\begin{equation}
G(\theta,\phi) = {4\pi r^2|\psi(r,\theta,\phi)|^2\over
{\rm Total~emitted~power}}.
\end{equation}
	
The absolute gain, as opposed to the relative gain, 
is important because it is the
quantity that indicates one's immunity to off-axis sources. If
the gains for all {\sl MAP's} frequency bands were the same
at some angle, each band would be
equally susceptible to a source at that angle. The gain is
always normalized so that  

\begin{equation}
\int G(\theta,\phi)d\Omega/4\pi =  
\int_{S'}{|\psi(x',y')|^2\over {\rm Total~emitted~power} }dx'dy' = 1, 
\end{equation} 
where the primed coordinates are for the aperture of the feed (or
optical element). 

The edge taper, $y_e$ is often useful in discussing the immunity of the 
optical system to sources in the sidelobes. In fact, from the
value of the field at the edge of an optic, one can compute the 
approximate shape of the sidelobe pattern. The pattern is then normalized
by the total power through the aperture. The edge taper is
given in dB as $y_e = 10\log(I_{edge}/I_{center})$
where $I$ is the intensity of the beam. For none of {\sl MAP}'s bands
is $y_e$ uniform around the edge; the 
largest (closest to $0~$dB) value is quoted. 

\section{Optical Design and Specification}
\label{sec:intro}

In designing \MAP, we considered a number of geometries
including simple offset parabolas and
three-reflector systems. In neither of these cases is the geometry of the feed 
placement conducive to having both inputs of a differential receiver
view the sky. With the back-to-back dual reflector arrangement we have chosen,
the feeds are centrally located and point in nearly opposite directions.
Additionally, the S/C moment of inertia is minimized for a large optic.

There are a number of dual-reflector telescope designs \cite{Schroeder,Love}.
For radio work the Cassegrain
(parabolic primary, hyperbolic secondary) and the Gregorian
(parabolic primary, elliptical secondary) are often used
\footnote{Though the Ritchey-Chretien (hyperbolic primary, hyperbolic
secondary, e.g., Hubble Space Telescope) and aplanatic
Gregorian (ellipsoidal primary, hyperbolic secondary) were considered,
the lack of well developed and tested offset designs did not fit with \MAP's
fast build schedule. Hanany \& Marrone (2002) give an up-to-date comparison of offset Gregorian designs.}. For some potential {\sl MAP} geometries, the scanning
properties of the Cassegrain system \cite{Ohm,RSG} were found superior 
to those of the Gregorian system. In other words, the
beam pattern from a feed placed a
fixed distance from the focus is more symmetric and has smaller 
near lobes. However, the offset Gregorian was chosen because the associated 
placement of the feeds was well suited to the differential receivers and
the FPA could occupy the space made available by
the position of the focus between the primary and secondary.
Additionally, for a given beam size, the Gregorian is more compact than
the Cassegrain \cite{Brown}. 

Dragone (1986, 1988) developed extremely low sidelobe Gregorian 
systems in which the
feed aperture is reimaged onto the primary. Such a system was used
for the ACME CMB telescope \cite{Meinhold}. 
Dragone's work was used as a
guideline but a number of factors complicated \MAP's design:
a) The feed apertures must be in roughly the same plane to avoid
being viewed by one another. b) The feed inputs must cover 
${\rm 20~cm \times 20~cm}$ to accommodate their large apertures. [The 
feed tails (receiver inputs) cover ${\rm 53~cm \times 53~cm}$ to make room for 
the microwave components; two W-band feed tails are separated by 19~cm; the two
Ka-band feed tails are separated by 45~cm.]
c) The secondary is in the near field of the feeds. 
For CMB telescopes, unlike the more familiar communications telescopes, 
beam efficiency is more important than 
aperture efficiency \cite{Rholfs}.   

To understand the beams, accurate computer codes are essential. 
The baseline \MAP design was done using code modified from
the work of Sletten (1988) in which the far field beam pattern
is computed from the square of the Fourier transform of the electric field
distribution in the aperture. This simple and fast code combined 
with parametric models of the
feeds and sidelobe response allowed rapid prototyping
of various geometries while simultaneously optimizing over
the combination of main beam size, contamination from the galactic pickup
through the sidelobes, and the sidelobe level at the position of the
Sun. All models followed the constraints for minimal cross-polar response 
\cite{TanMiz,Mizu76}. The parameters of resulting telescope are given in 
Table~\ref{tab:refl_params}.

\begin{figure*}[tb]
\epsscale{1.4}
\plotone{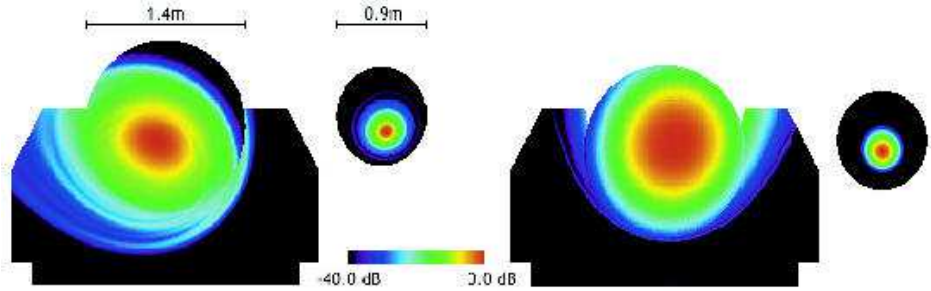}
\caption{The current density on the primary and secondary reflectors
for K-band (22 GHz) on the left and W-band (90 GHz) on the right.
The 1.4~m diameter disk shows the projection of the primary as 
seen from the optical axis. The 2.7~m by 1.6~m rectangle with the 
modified corners shows the projection of the radiator
panels. (The net emitting area of the panels is ${\rm 5.45~m^2}$.)
As the far field approximates the square of the Fourier
transform of the current density, one can see that 
the K-band beam will not be as symmetric as the W-band beam.
The reflectors are in the near field of the telescope system so the
apparent fraction of the beam spilled to the radiators is not indicative
of the solid angle on the sky.}
\label{fig:curd}
\end{figure*}

\begin{figure*}
\epsscale{1.5}
\plotone{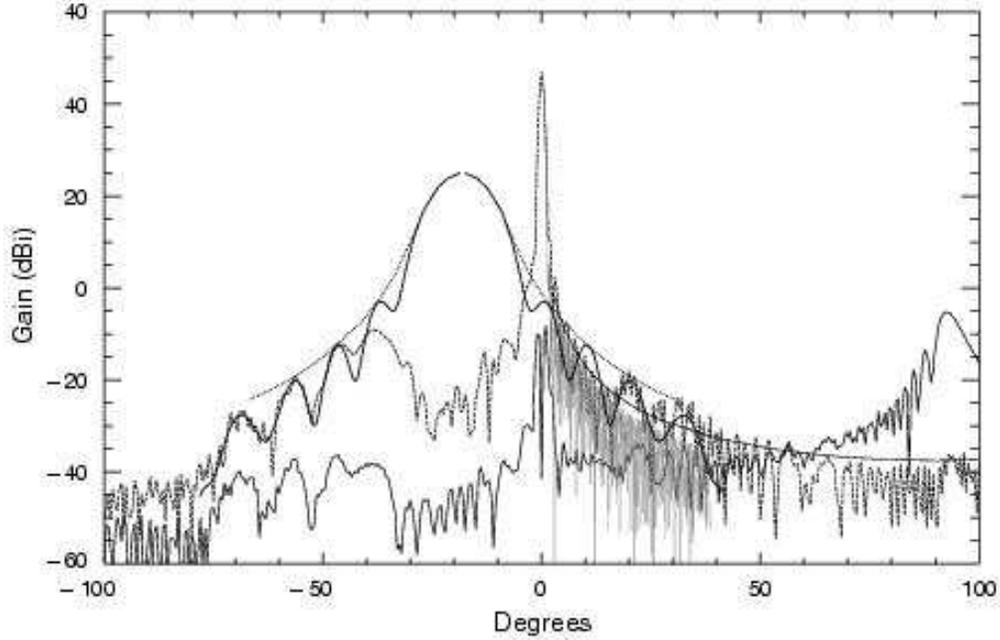}
\caption{The optical response of a prototype design with a K-band feed
at the focus of the telescope. The y-axis is the
gain; the x-axis shows the angular distance 
from the primary optical axis. Negative numbers correspond to lines of
sight below the optical axis as shown in Figure~\ref{fig:sideview}. 
The dashed line is the full model computed by the {\tt DADRA} code for
left circular polarization. The
broad solid line that peaks at $-18\deg$ would be 
the response of the feed if it
were not blocked by the secondary. Note that where the 
secondary does not intercept the feed pattern, $-40>\theta>15\deg$,
the full pattern is dominated by the feed pattern.
In other words, spill of the feed past the secondary dominates 
the optical response. The smooth dotted curve that envelopes the feed
response is the parameterized model discussed in Section~\ref{sec:feeds}.
The smooth line extending from the peak to $\theta=100\deg$ is the
parameterized sidelobe response discussed in Section~\ref{sec:sidelobes}. 
The light curve for $0\deg<\theta<40\deg$ is from the aperture 
integration code.
The curve that peaks at $\theta=90\deg$ is the response to radiation that 
has reflected once off the bottom of the secondary and into the feed 
and thus is right circularly polarized. This specific configuration 
was not built.} 
\label{fig:k_composite}
\end{figure*}

The {\tt DADRA} ``physical optics'' computer code \cite{YRS} was 
used for the detailed design of \MAP. From a spherical wave expansion 
of the field from a feed, the code determines the surface current 
on the secondary, ${\bf J}( r) = 2\hat {\bf n}\times{\bf H}_{inc}(r)$.
 From this current,
it computes the fields incident on the primary and thus the currents there.
These currents are particularly useful for understanding the interaction
of the optics with the S/C components. Examples are shown in 
Figure~\ref{fig:curd} for the lowest and highest frequency bands.
The resulting beam is a sum of the
fields from the feed and the currents on the primary and secondary
reflector surfaces as shown in Figure~\ref{fig:k_composite}.
The method takes into account the vector nature of the fields and the
exact geometry of the reflectors and the feeds. In principle, the currents 
can be unphysical near the reflector edges.  In practice,
because of the low edge taper, this approximation does not introduce
significant errors.
The code is excellent though its limitations are that
a) only two reflections are considered, b) the possible interference of the 
feeds with the radiation from the reflectors and with each other is 
not accounted for, and c) the interactions with the structure are not accounted
for.  These interactions must be determined ``by hand'' 
and thus measurements of the assembled system are essential. 
 
\subsection{Flight Design}

Once the baseline Gregorian design was set, the surfaces 
were ``shaped'' to optimize the symmetry and size of the beams
\cite{YRS,Galindo92}. The resulting reflectors differ from the pure Gregorian by
$\approx~1~$cm in regions near the perimeter
\footnote{In retrospect we should not 
have shaped the reflectors. It added time to the manufacturing 
process and the cool down distortions of the primary reflectors
negated its benefits.}. For simple calculations, we use
the best fit parameters shown in Table~\ref{tab:refl_params}. 

\begin{table*}[t]
\caption{\small Parameters of One Telescope}
\small{
\vbox{
\tabskip 1em plus 2em minus .5em
\halign to \hsize {#\hfil &\hfil#\hfil &\hfil#\hfil &\hfil#\hfil  \cr
\noalign{\smallskip\hrule\smallskip\hrule\smallskip}
Quantity  & Base Design & Best fit & Equiv parabola\cr
Focal length, $f_p$   (cm)           & 90 & 90     & 206 \cr
Primary projected radius, $r_p$ (cm)  & 70 & 70     &70\cr
Offset parameter, $y_c$  (cm)         & 105 & 104.03 & \dots  \cr
Interfocal distance, $2c$  (cm)       & 45 & 43.58  & \dots  \cr
Secondary eccentricity, $e$      & 0.45 & 0.4215 & \dots  \cr
$\beta$    (deg)                      & 12.06 & 13.77  &  \dots \cr
$\alpha$   (deg)                      &-31.12  & -33.06 &  \dots \cr
\noalign{\smallskip\hrule}
}}}
\small{Cassegrain and Gregorian telescopes are defined by 5 parameters
when they satisfy the minimum cross-polarization condition. 
These parameters are shown
in Figure~\ref{fig:sideview}. Column 1 is the nominal
conic design. Column 2 is the best fit to the shaped design with 5
free parameters. Column 3 shows the equivalent parabola \cite{Rausch} for 
the best fit
design described by the top five parameters. When the best fit
design is constrained to follow the minimum cross-polarization condition,  
$\beta=14.32^{\circ}$ and $\alpha=-34.32^{\circ}$. The surface coating is 
vacuum deposited aluminum as discussed in section 2.8.}
\label{tab:refl_params}
\end{table*} 

The shaped system does not have a sharp focus between the
primary and secondary but
one may still characterize its response in broad terms.
The plate scale, how far one moves laterally in the focal plane to
move a degree on the sky, for the shaped system is 4.44 cm/deg.
This suggests an effective focal length of 250 cm, somewhat longer
than that found from the equivalent parabola for the best fit model. 
The speed of the system, $f_p/D$ is f/1.8. The computer files containing
the geometry are available upon request. 

\subsection{Manufacture and Alignment of Optics}

The TRS and REU were built by Programmed Composites Inc. (PCI) to a 
specification \cite{Jackson94}. The structure that 
holds the optics is made of 5 cm by 5 cm ``box beams'' 
of 0.76~mm thick XN70/M46J composite material and
has a mass of 23 kg. The reflectors are made of 0.025 cm thick XN70
spread fabric cloth face sheets\footnote{At room temperature, the
material has a resistance of $1-2~\Omega$ as measured diagonally
across the reflector. The thermal conductivity at 290~K is $\approx0.91$~W/cmK,
nearly half that of aluminum.} over a 0.635 cm thick DuPont KOREX honeycomb
core. The combination
of materials was chosen based on PCI models that predicted 
a negligible net coefficient of thermal expansion between 70 and 300~K.
The mass of one primary, with the backing structure, is 5 kg. The
mass of one secondary is 1.54 kg. The radiator panels are made of 1100
series H14 aluminum over a 5.08 cm aluminum honeycomb core and each has
a mass of 8 kg. They are painted with NS43G/Hincom white paint to
minimize their solar absorptance, ensure a conducting surface, and
maximize their infrared emissivity. The full TRS, with harnesses and
thermal blankets, has a mass of 70 kg.

The specifications were set to meet the science goals and to easily mesh
with known tolerances in
the manufacturing process to keep costs down. The surface {\it rms}
deviation at 70~K from the ideal shape over the whole reflector 
was specified to be $<0.0076~$cm or $\approx\lambda/40$, and is discussed in
more detail below. The reflectors, when treated as rigid bodies
satisfying the surface {\it rms} criteria and when positioned on the TRS
structure and cooled to 70K, were specified to be within 0.038~cm of the design
position. This specification includes all rotations and translations and accounts
for the effects of moisture desorption, gravity relief, and cooldown
from room temperature. 
The on-orbit predictions are given in Table~\ref{tab:coord}. 

\begin{table*}[t]
\caption{\small S/C Coordinates for A-side}
\small{ \vbox{
\tabskip 1em plus 2em minus .5em
\halign to \hsize {
     #\hfil & 
\hfil#\hfil &
\hfil#\hfil &
\hfil#\hfil \cr
\noalign{\hrule\vskip1pt\hrule\smallskip}
    Object    &
    $x_s$ (cm)& 
    $y_s$ (cm)& 
    $z_s$ (cm)\cr
    Focus    &
     0 & 
     65.0 & 
     -166.0 \cr
    Top of A Primary    &
    0.045 & 
    35.99 & 
    -351.66 \cr
    Bottom of A Primary    &
    0.040 & 
    7.58 & 
    -192.75 \cr
 $+x$ Boundary of A Primary    &
    70.05 & 
    21.77 & 
    -272.30 \cr
    Top of A Secondary    &
     0.031 & 
    123.31  & 
    -215.85 \cr
    Bottom of A Secondary    &
    -0.011 & 
    102.20 & 
    -136.00 \cr
 $+x$ Boundary of A Secondary    &
    39.12 & 
    36.73 & 
    -176.36 \cr
\noalign{\vskip1pt\hrule\vskip1pt\hrule\smallskip}
}}}
\label{tab:coord}
\end{table*}

The optics were built to have no adjustments. They were designed to be 
in focus at 70~K and so were deliberately though insignificantly
out of focus for all testing at 290~K. A full STOP (Structural
Thermal OPtical) performance analysis of the optical design
was performed using the W-band
beam pattern because it is the most sensitive to changes in the
position. The analysis includes changes in the position and orientation
of the optics and feeds as they cool. The worst case displacements 
upon cooling lead to 
a $0.1^{\circ}$ shift in beam elevation and a $0.14^{\circ}$ shift
in azimuth, dominated by orientation changes in the primary.
These are not significant from a radiometric point of view.
In addition, all final pointing and beam information is determined 
from in-flight observations.  
Table~\ref{tab:fpa_beams} shows the placement of the feeds
along with the predicted beam positions on the
sky. Tables~\ref{tab:refl_params}, \ref{tab:coord} and
\ref{tab:fpa_beams} completely specify the geometry for the conic
approximation.

In addition to the usual metrology tools, photogrammetry and 
laser tracking were found to be particularly useful. Photogrammetry was used 
to determine the change in the shape of the optics upon cooling 
to 70~K. The laser tracker allowed the rapid digitization of thousands
of points on the reflectors which was useful for surface fitting and
measuring the surface deformations. 
 
\begin{table*}[t]
\caption{\small Design Positions of Main Beams}
\small{
\vbox{
\tabskip 1em plus 2em minus .5em
\halign to \hsize {\hfil#\hfil & \hfil#\hfil &\hfil#\hfil 
&\hfil#\hfil &\hfil#\hfil &\hfil#\hfil \cr
\noalign{\smallskip\hrule\smallskip\hrule\smallskip}
Beam &Aperture & OMT to Apt vector & Direction on sky &  P1 & P2 \cr
K1A 
& \scriptsize{(9.49,73.21,-177.06)} 
& \scriptsize{(-0.2303,0.9438,0.2370)}
& \scriptsize{(0.0404, 0.9246, -0.3788)}
& \scriptsize{(0.6927, -0.2991, -0.6562)}
& \scriptsize{(-0.7200,-0.2358,-0.6526)} \cr
K1B 
& \scriptsize{(9.47,-73.26,-177.08)} 
& \scriptsize{(-0.2301,-0.9438,0.2372)}
& \scriptsize{(0.0407,-0.9240, -0.3803)}
& \scriptsize{(0.6951,0.2996,-0.6535)}
& \scriptsize{(-0.7178,0.2377,-0.6545)}\cr
Ka1A 
& \scriptsize{(-8.66,73.35,-176.24)} 
& \scriptsize{(0.2438,0.9367,0.2515)}
& \scriptsize{(-0.0377, 0.9258, -0.3762)}
& \scriptsize{(-0.6964, -0.2943, -0.6545)}
& \scriptsize{(0.7167, -0.2373, -0.6558)}\cr
Ka1B 
& \scriptsize{(-8.64,-73.42,-176.24)} 
& \scriptsize{(0.2434,-0.9371,0.2501)}
& \scriptsize{(-0.0374, -0.9251, -0.3778)}
& \scriptsize{(-0.6938, 0.2961,-0.6564)}
& \scriptsize{(0.7191, 0.2376, -0.6530)}\cr
Q1A 
& \scriptsize{(-7.20,71.93,-158.79)} 
& \scriptsize{(0.2146,0.9527,-0.2152)}
& \scriptsize{(-0.0314, 0.9523, -0.3034)}
& \scriptsize{(0.7119, -0.1918, -0.6756)}
& \scriptsize{(-0.7015, -0.2373, -0.6720)} \cr
Q1B 
& \scriptsize{(7.19,-71.99,-158.78)} 
& \scriptsize{(0.2155,-0.9524,-0.2155)}
& \scriptsize{(-0.0319, -0.9522,-0.3037)}
& \scriptsize{(0.7147, 0.1907, -0.6729})
& \scriptsize{(-0.6987, 0.2385, -0.6745})\cr
Q2A 
& \scriptsize{(7.20,71.93,-158.79)} 
& \scriptsize{(-0.2150,0.9527,-0.2148)}
& \scriptsize{(0.0328, 0.9523, -0.3034)}
& \scriptsize{(-0.7147, -0.1899, -0.6731)}
& \scriptsize{(0.6986, -0.2389, -0.6744)}\cr
Q2B 
& \scriptsize{(7.19,-71.98,-158.77)} 
& \scriptsize{(-0.2153,-0.9523,-0.2161)}
& \scriptsize{(0.0323, -0.9522, -0.3037)}
& \scriptsize{(-0.7121, 0.1912, -0.6755)}
& \scriptsize{(0.7013, 0.2381, -0.6719)} \cr
V1A 
& \scriptsize{(-8.00,71.00,-167.40)} 
& \scriptsize{(0.2076,0.9782,0.0082)}
& \scriptsize{(-0.0328, 0.9416, -0.3352)}
& \scriptsize{(-0.6976, -0.2617, -0.6669)} 
& \scriptsize{(0.7157, -0.2119, -0.6659)}\cr
V1B 
& \scriptsize{(-7.93,-71.07,-167.37)} 
& \scriptsize{(0.2072,-0.9783,0.0071)}
& \scriptsize{(-0.0333, -0.9415, -0.3354)}
& \scriptsize{(-0.6994, 0.2617, -0.6651)}
& \scriptsize{(0.7140, 0.2125, -0.6672)}\cr
V2A 
& \scriptsize{(8.00,71.00,-167.40)} 
& \scriptsize{(-0.2077,0.9782,0.0076)}
& \scriptsize{(0.0335, 0.9416, -0.3352)}
& \scriptsize{(0.6986, -0.2620, -0.6660)}
& \scriptsize{(-0.7149, -0.2118, -0.6664)}\cr
V2B 
& \scriptsize{(7.92,-71.07,-167.37)} 
& \scriptsize{(-0.2073,-0.9783,0.0071)}
& \scriptsize{(0.0331, -0.9415, -0.3354)}
& \scriptsize{(0.6974, 0.2621, -0.6670)}
& \scriptsize{(-0.7160, 0.2119, -0.6652)}\cr
W1A 
& \scriptsize{(-2.40,70.02,-170.00)} 
& \scriptsize{(0.0432,0.9978,0.0501)}
& \scriptsize{(-0.0087, 0.9396, -0.3422)}
& \scriptsize{(0.7098, -0.2352, -0.6639)}
& \scriptsize{(-0.7043, -0.2487, -0.6649)}\cr
W1B 
& \scriptsize{(-2.40,-70.15,-170.03)} 
& \scriptsize{(0.0428,-0.9979,0.0496)}
& \scriptsize{(-0.0095, -0.9394, -0.3428)}
& \scriptsize{(0.7100, 0.2351, -0.6638)}
& \scriptsize{(-0.7041, 0.2497, -0.6647)}\cr
W2A 
& \scriptsize{(-2.40,69.98,-165.20)} 
& \scriptsize{(0.0364,0.9989,-0.0290)}
& \scriptsize{(-0.0091, 0.9459, -0.3245)}
& \scriptsize{(-0.7058, -0.2359, -0.6679)} 
& \scriptsize{(0.7083, -0.2229, -0.6697)}\cr
W2B 
& \scriptsize{(-2.46,-70.11,-165.21)} 
& \scriptsize{(0.0356,-0.9989,-0.0302)}
& \scriptsize{(-0.0095, -0.9458, -0.3247)}
& \scriptsize{(-0.7056, 0.2365, -0.6680)}
& \scriptsize{(0.7085, 0.2228, -0.6696)} \cr
W3A 
& \scriptsize{(2.40,69.98,-165.20)} 
& \scriptsize{(-0.0351,0.9990,-0.0289)}
& \scriptsize{(0.0101, 0.9457, -0.3248)}
& \scriptsize{(0.7058, -0.2368, -0.6677)}
& \scriptsize{(-0.7083, -0.2225, -0.6699)}\cr
W3B 
& \scriptsize{(2.45,-70.08,-165.19)} 
& \scriptsize{(-0.0368,-0.9989,-0.0296)}
& \scriptsize{(0.0093, -0.9458, -0.3247)}
& \scriptsize{(0.7049, 0.2365, -0.6687)}
& \scriptsize{(-0.7092, 0.2227, -0.6689)}\cr
W4A 
& \scriptsize{(2.40,70.02,-170.00)} 
& \scriptsize{(-0.0428,0.9978,0.0500)}
& \scriptsize{(0.0101, 0.9394, -0.3426)}
& \scriptsize{(-0.7095, -0.2347, -0.6645)}
& \scriptsize{(0.7046, -0.2498, -0.6642)}\cr
W4B 
& \scriptsize{(2.39,-70.18,-170.02)} 
& \scriptsize{(-0.0436,-0.9978,0.0496)}
& \scriptsize{(0.0093, -0.9394, -0.3428)}
& \scriptsize{(-0.7102, 0.2351, -0.6636)}
& \scriptsize{(0.7039, 0.2497, -0.6650)} \cr
\noalign{\smallskip\hrule}
}}}
\small{These values and unit vectors are in S/C coordinates and thus can be
directly compared to Table~\ref{tab:coord}. The final beam position will be
determined in flight. The ``aperture'' is the coordinate in the focal plane.
``OMT to Apt'' is the unit vector along the symmetry axis of the feed. 
``Direction on sky'' is the unit vector along the optical axis. 
The polarization directions, P1 and P2, correspond to the
maximum electric field for the main (or axial) and side (or radial) OMT 
ports respectively.  The ``1'' or ``2'' is the last number in a 
radiometer designation. A DA differences two polarizations of
similar orientation on the sky as can be seen by comparing K11A, polarization 
direction 1 (P1) on the A side, and K11B, the matching input to the 
DA on the B side. } 
\label{tab:fpa_beams}
\end{table*}

\subsection{Feeds}
\label{sec:feeds}

The inputs of the differential microwave receivers are coupled 
to free space with corrugated microwave feeds \cite{Barnes02}. 
The initial designs followed well known 
principles \cite{Thomas,Clarricoats} though the final groove dimensions
were optimized \cite{YRS}. Corrugated feeds were chosen because
a) their patterns are accurately computable and symmetric;  b) they 
have low loss; and c) they have low slidelobes.
Table~\ref{tab:feeds_omt} summarizes the features of the {\sl MAP}
feeds.  

At the base of each feed is an orthomode transducer (OMT),
the microwave analog of a polarizing beam splitter.
The two rectangular waveguide outputs
of the OMT, the ``main'' port and the ``side'' port,  
carry the two constituent polarizations to the inputs 
of separate differencing assemblies \cite{Jarosik03}. 
As the polarization cannot be 
determined in flight, it is completely characterized on the ground.

The beam from each feed is diffraction limited so that 
the illumination patterns on the secondary and primary are a function 
of frequency. Because the secondaries are in the near
field of the feeds, $d<2d_{fapt}^2/\lambda$,  the phase center concept
is not useful for detailed predictions. A full electromagnetic code such as
{\tt CCRHRN} \cite{YRS} is essential. However, 
for the top level parameterization of {\sl MAP} the feeds were modeled as open
corrugated wave guides with $B_n(\theta)=[x_1^2J_0(v)/(x_1^2-v^2)]^2$
where $v=kr_1\sin(\theta)$, $x_1\approx2.405$, $k=2\pi/\lambda$, and
$r_1$ is the radius of the waveguide \cite{Clarricoats}. At large 
angles, the envelope 
$J_0(v)\approx\sqrt{2/v\pi}$  is shown in Figure~\ref{fig:k_composite}
for a K-band feed. 

\begin{table*}[ht]
\caption{\small Feeds and OMTs}
\small{ \vbox{
\tabskip 1em plus 2em minus .5em
\halign to \hsize {
     #\hfil & 
\hfil#\hfil &
\hfil#\hfil &
\hfil#\hfil &
\hfil#\hfil &
\hfil#\hfil \cr
\noalign{\hrule\vskip1pt\hrule\smallskip}
    Band   &
    K & 
    Ka & 
    Q & 
    V & 
    W \cr
    Center frequency (GHz) & 
    23  & 
    33  & 
    41  & 
    61  & 
    93  \cr
  $\theta_{FWHM}^{\rm feed}$ (deg) &
    $8.8^{\circ}$ & 
    $8.3^{\circ}$ &
    $7.0^{\circ}$  &
    $8.0^{\circ}$  &
    $8.4^{\circ}$ \cr
  $G_m^{\rm feed}$  (dBi) &
    26.79 & 
    27.23 &
    28.77  &
    27.28  &
    26.40 \cr
  $d_{fapt}$~(cm)  &
    10.94 & 
    8.99 &
    8.99  &
    5.99  &
    3.99 \cr
  $Crosspol^{omt}$ (dB) &
    -40 & 
    -30 &
    -30 &
    -27 &
    -25 \cr
\noalign{\vskip1pt\hrule\vskip1pt\hrule\smallskip}
}}}
\small{See also Jarosik {\it et al.} 2003. The maximum gain, $G_m$, is 
for a lossless feed.}
\label{tab:feeds_omt}
\end{table*}

\subsection{Main Beams}

The main beam width is determined by the size of the
primary mirror, the edge taper (Table~\ref{tab:sidelobes}) and to a lesser
degree, the beam profiling.
As discussed in Section~\ref{sec:sidelobes}, the
edge taper is $\approx -20~$dB, except in K-band.
With a 1.4~m projected primary diameter, the beam width is
$\theta_{fwhm} \approx 0.5^{\circ}(40~{\rm GHz}/\nu)$.
Because a diffraction limited feed illuminates the primaries and secondaries
with a low edge taper, the beam width 
is a relatively weak function of frequency within a frequency band
as shown in Table~\ref{tab:kw_beams}.

The number of feeds and radiometers
in each frequency band is chosen so that each band has roughly
equal sensitivity per unit solid angle to celestial microwave radiation. 
Because the noise
temperature of HEMT amplifiers scales approximately with frequency
\cite{Posp95},
{\sl MAP} has 1 feed in K-band and Ka-band, 2 feeds in Q-band and V-band,
and 4 in W-band.

The desire for broad frequency coverage and multiple channels requires the
maximal use of the telescope focal plane. 
The layout of the two back-to-back telescopes shown in Figure~1, and 
the need for a compact enclosure for all the differential radiometers 
places difficult
constraints on the geometry of the feeds, the radiometers, and the focal-plane
arrangement. In particular, the need for roughly equal total feed length, the
requirement for minimum cross-talk and obscuration between feeds 
in the focal plane, and the
placement of the OMTs and cold amplifiers lead to feed-telescope solutions which
required evaluation with full diffraction calculations for optimizing the
configuration. For example,
the K-band feed is profiled to reduce its length by a factor of $\approx 50$\%
from its
nominal geometry. The K-band feed is also shifted along its axis
by 15 cm toward the secondary from its optimal position. Such large
departures from usual practices
are acceptable if the design can be evaluated and the beam profiles
and solid angles can be measured to $\approx0.5$\% accuracy in flight.

In general, a controlled loss of axial symmetry of the beam point spread 
function was balanced against other geometrical constraints as the
whole system was considered simultaneously. 
The solution for the reflectors results in an uninverted image of the sky on a
slightly curved focal plane. The geometric image quality degrades with
increasing distance from the optics axis. Within a $2^{\circ}$ 
diameter centered
on the optical axis, 95\% of the incident power falls within a 0.4~cm diameter
disk. Further out, within a $4^{\circ}$ diameter region the 95\% disk 
is 1~cm in diameter.
 
\begin{table*}[t]
\caption{Predicted ideal K-band and W-band main beams}
\small{
\vbox{
\tabskip 1em plus 2em minus .5em
\halign to \hsize {
     #\hfil & 
\hfil#\hfil &
\hfil#\hfil &
\hfil#\hfil &
\hfil#\hfil \cr
\noalign{\hrule\vskip1pt\hrule\smallskip}
K-band Frequency (GHz) &
20 &
22 &
25 \cr
\noalign{\smallskip\hrule\smallskip}
Feed 3 dB width (deg)    &
  10.0  & 
  8.9   & 
  7.7   & 
  $\dots$ \cr
Vertical 3~dB width\tablenotemark{a} (deg)    &
  0.969  & 
  0.882   & 
  0.787    &
  $\dots$   \cr 

Horizontal 3~dB width (deg)    &
  0.798  & 
  0.721   & 
  0.637    &
  $\dots$   \cr 
$G_m$ (dBi)  &
  46.30  & 
  46.97   & 
  47.79    &
  $\dots$   \cr 
$\Omega_A$ (sr) &
  $2.82\times10^{-4}$  & 
  $2.44\times10^{-4}$   & 
  $2.03\times10^{-4}$    &
  $\dots$  \cr
$\sqrt{\Omega}/\lambda$ (cm$^{-1}$)\tablenotemark{b}&
  0.0112  & 
  0.0115  & 
  0.0119  &
  $\dots$  \cr
$A_{eff}$ (m$^2$)\tablenotemark{c}  &
  0.80  & 
  0.76  & 
  0.71  &
  $\dots$  \cr
Primary edge taper (dB)    &
  $-12.8$  & 
  $-13.1$  & 
  $-14.7$  &
  $\dots$  \cr
\noalign{\vskip1pt\hrule\smallskip}
W-band Frequency (GHz) &
82 &
90 &
98 &
106 \cr
\noalign{\smallskip\hrule\smallskip}
Feed 3 dB width$^{a}$ (deg)    &
  27.9  & 
  25.0   & 
  $\dots$   & 
  21.4   \cr
$+45^{\circ}$ slice 3 dB beam width (deg)    &
0.209 & 
0.201 & 
0.198 & 
0.194 \cr
$-45^{\circ}$ slice 3 dB beam width (deg)    &
0.199  & 
0.190  & 
0.184  & 
0.181  \cr
$G_m$ (dBi)    &
59.40   & 
59.75   & 
59.95   & 
60.09   \cr
$\Omega_A$ (sr)  &
  $1.43\times10^{-5}$  & 
  $1.33\times10^{-5}$  & 
  $1.27\times10^{-5}$  &
  $1.23\times10^{-5}$    \cr
$\sqrt{\Omega}/\lambda$ (cm$^{-1}$)  &
  0.0103  & 
  0.0109  & 
  0.0116   & 
  0.0124   \cr
$A_{eff}$ (m$^2$)  &
  0.94  & 
  0.83  & 
  0.74   & 
  0.65    \cr
Primary edge taper  (dB)    &
$-17.4$  & 
$-21.0$  & 
$-24.8$  & 
$-26.5$  \cr
\noalign{\vskip1pt\hrule\vskip1pt\hrule\smallskip}
}}}
\tablenotetext{a}{As shown in Figure~\ref{fig:beam_maps}, the orientation of
the long axis of the ellipsoidal beam shape is approximately vertical.
For the W-band beams, the $\pm 45^{\circ}$ $\theta_{fwhm}$ are shown as
they represent the largest departures from circularity.}
\tablenotetext{b}{For a diffraction limited beam feeding
diffraction limited optics with a low edge taper, the gain is independent of
frequency and so $\sqrt{\Omega}/\lambda$ is constant.
Departures from this are due to the fact that the optical elements are 
not in the far field of each other and the feeds and the finite edge taper.}
\tablenotetext{c}{The physical area of the primary is 1.76~m$^2$ so that the
aperture efficiency is $\approx 0.5$}
\label{tab:kw_beams}
\end{table*}

\subsection{Polarization}

The optical system is designed to minimize the cross-polarization.
The OMTs attached to the azimuthally symmetric feed defines
the polarization direction. Each OMT is oriented so that the polarization
directions accepted by the feeds are $\pm 45^{\circ}$ 
with respect to the $yz$ symmetry plane 
of the satellite. The unit direction vectors on the sky are given in 
Table~\ref{tab:fpa_beams}.
The $45^{\circ}$ angle ensures that the two linear polarizations 
in one feed nearly symmetrically illuminate the primary and secondary.
Consequently, the beam patterns for both polarizations 
are nearly identical. 

There are some subtleties in specifying the polarization angle.
For example, the average of
the polarization angle within a $-3$~dB contour of the main beam
does not equal
the polarization direction at the beam maximum. It can differ by of order a
degree. {\sl MAP's} angles were set to $\pm 45^{\circ}$ at the beam maximum.
Even though the corresponding orientation of the OMT was found using the 
full diffraction calculation, the same orientation is obtained with 
purely geometrically considerations. 

The {\sl MAP} polarization angle was measured in the NASA/GSFC 
beam mapping facility.
A polarized source was rotated to minimize the average signal across all
frequencies in the band. The polarization angle was then taken to 
be $90^{\circ}$ from that angle. The measured polarization minimum does 
not occur at the same angle across the band and so a best overall
minimum angle was approximated. The uncertainty of
the radiometric measurement is $\pm 1^{\circ}$. Across all bands and all
feeds, the scatter in polarization angle is $\pm 2^{\circ}$. 
The uncertainty in the polarization angle based
on the metrology of the OMT orientation is $\pm 0.2^{\circ}$.
We use $\pm 1^{\circ}$ as the formal error.  The
resulting maximum misidentification of power due to the rotational alignment
uncertainty is $\approx 1 - \cos^2(1^{\circ})\approx 0.0003$.

The cross-polar leakage is not zero but is sufficiently small to allow the
measurement of the CMB
polarization to the limits of the detector noise.  In K and Ka bands, the cross
polar contribution to the co-polar beam is $-25$~dB and $-27$~dB respectively,
and is dominated by the reflector geometry in combination with the 
feed placement. In Q, V, and W bands, the cross-polar contributions to 
the co-polar response are $-24$~dB , $-25$~dB, and $-22$~dB respectively 
and arise from a combination of imperfect OMTs,
as shown in Table~\ref{tab:feeds_omt}, and reflector geometry.   

\subsection{Surface Shape}
\label{sec:surface}
 
The smoothness of the optical surface and the placement of the feeds
determine the quality of the beams. The measured beam profiles,
especially in W-band, do not precisely match the predictions for ideal 
reflectors. However, they are in
excellent agreement with the computations based on the measured
distortions of the reflector surfaces. 

The characteristics of the profiles are also in good
agreement with the Ruze (1966) model which
accounts for the axial loss of gain and the pattern
degradation as a function of the reflector surface 
{\sl rms} error, $\sigma_z$, and
the spatial correlation length of the distortions, $l_c$. 
For lossless Gaussian shaped distortions, which sufficiently describe
the surfaces,

\begin{eqnarray} 
\lefteqn{ G(\theta,\phi) = G_0(\theta,\phi) e^{-\overline{\rho^2}} +}
\hspace{2.8truein}\nonumber \\
\left( \frac{2 \pi l_c}{\lambda} \right)^2 e^{-\overline{\rho^2}}
\sum_{n=1}^{\infty} \frac{\overline{\rho^2}^n}{n \times n!}
e^{-\left(\pi l_c \sin(\theta) / \lambda \right)^2/n}
\label{E:Gauss_gain}
\end{eqnarray}
where
$\lambda$ is the wavelength, $G_0(\theta,\phi)$ is the
ideal undistorted prediction, 
$\overline{\rho^2}$ is the variance of the phase error, equal to 
$k^2 {\sigma_z}^2$, and $l_c$ is the correlation length. 
The number of distorted ``lumps,'' $\approx ( D/2 l_c )^2 \approx  50$,
is large enough to satisfy the statistical assumptions behind  Ruze's model.

The first term in equation~\ref{E:Gauss_gain} shows that the reduction in
forward gain from the undistorted reflector is determined
by $\sigma_z$ and is independent of $l_c$. The second term,
the ``Ruze pattern,'' is a
function of $\theta$, is independent of the undistorted pattern, and
is determined by $l_c$ and  $\overline{\rho^2}$.  The shoulder 
of the Ruze beam 
is mostly determined by $l_c$. Increasing the $\sigma_z$ from zero while
keeping $l_c$ constant lowers the forward gain and raises
the Ruze beam, without changing the shoulder. Increasing $l_c$ 
from zero narrows the Ruze pattern.

The surface shape was specified to achieve close to 
ideal performance. Ground testing revealed that the 
primary mirror shape met the specification
at room temperature but would not
at the second Earth-Sun Lagrange point
\footnote{Because {\sl MAP} exceeded the
higher-level specified resolution of $0.3^{\circ}$,
no corrective action was taken so as to protect {\sl MAP's}
schedule and budget caps.}, L$_2$,
from where {\sl MAP} observes \cite{Bennett03}.
At room temperature, the surface parameters are 
computed from measurements of the surface made with a 
laser tracker. The cold shape of the surface was determined through 
photogrammetry of 35 targets affixed to a 90~K primary 
and then extrapolated to 70~K, the prediction for L$_2$.
The results are given in Table~\ref{tab:surf_parms}.
The departure from the ideal shape is rooted in errors in the estimates 
of the thermal coefficients of expansion of some of the materials.
Even though the diameter of the secondary is two thirds that of the primary,
it is less susceptible to cool-down distortions. This is attributed
to its cylindrically symmetric mount and more extensive backing structure
as opposed to the ``goal post'' mount for the primaries. 

\begin{table*}[t]
\caption{Reflector Surface Shape Parameters}
\small{ \vbox{
\tabskip 1em plus 2em minus .5em
\halign to \hsize {
     #\hfil & 
\hfil#\hfil &
\hfil#\hfil &
\hfil#\hfil &
\hfil#\hfil &
\hfil#\hfil &
\hfil#\hfil &
\hfil#\hfil &
\hfil#\hfil \cr
\noalign{\hrule\vskip1pt\hrule\smallskip}
Reflector & Spec $\sigma_z$\tablenotemark{a}
          & $\sigma^{290K}_z$ (cm)
          & $\sigma_z^{70~K}$ (cm)
          & $l_c$ (cm) 
          & $e^{-\overline{\rho^2}}$ warm \tablenotemark{b}
          & $e^{-\overline{\rho^2}}$ cold \tablenotemark{b} 
          & $k l_c$ \cr
Primary	  P3 (A side) &	0.0076  & 0.0071 & 0.023   & 9.3  & 0.989    
                      & 0.826  & 175 \cr
Primary   P2R (B side) &  0.0076  & 0.0071 & 0.024   & 11.4 & 0.987    
                      & 0.812  & 213 \cr
Secondary S4 (A side) &	0.0076  & 0.0071 &$\cdots$ &10.1 & 0.988 
                      & $\cdots$  & 191 \cr
Secondary S5 (B side) &	0.0076  & 0.0076 &$\cdots$ &9.7 & 0.991 
                      & $\cdots$  & 183 \cr
\noalign{\vskip1pt\hrule\vskip1pt\hrule\smallskip}
}}}
\tablenotetext{a}{The surface rms, $\sigma_s$, was specified
as a function of
radius to mesh with manufacturing processes. For the primary 
$\sigma_s<0.0038~$cm for $r<25~$cm and $\sigma_s<0.0051$~cm for $r<50~$cm.
For the secondary $\sigma_s<0.0038~$cm for $r<15~$cm and
$\sigma_s<0.0051$~cm for $r<35~$cm. For equation 6, we convert
the surface distortions, $\sigma_s$, to distortions projected along
the primary optical axis, $\sigma_z$.}
\tablenotetext{b}{The reduction in forward gain due to scattering for
the ambient temperature and on-orbit cases.  The warm reflectors meet
the specification 
and scatter just slightly over 1\% away from the forward direction. 
The cold reflectors scatter out of order 20\%, or 1~dB, out of the main
beam and into the near sidelobes as shown in Figure~\ref{fig:srf_gain}.}
\label{tab:surf_parms}
\end{table*} 

\begin{figure*}
\epsscale{1.6}
\plotone{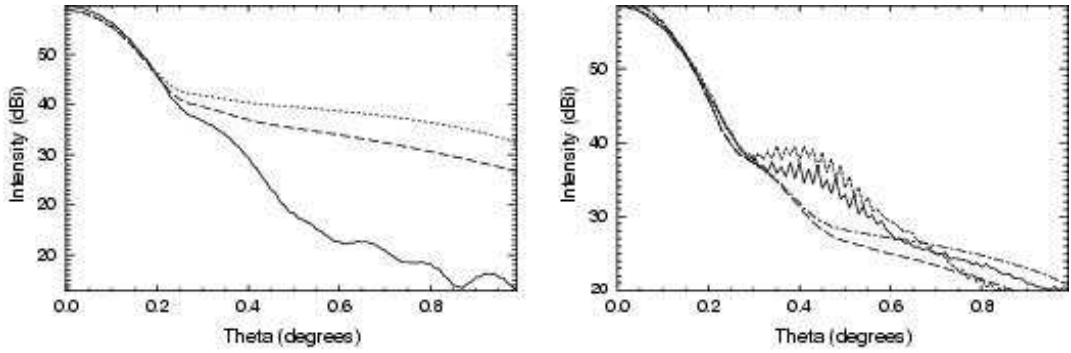}
\caption{{\it Left:} Beam profiles at 90 GHz from a distorted surface
based on Gaussian 
deformations  (dashed line, eq.~\ref{E:Gauss_gain}) and hat box 
distortion (dash-dot line, Ruze (1966)). The azimuthally averaged
measured warm profile 
(solid line) is shown with the {\tt DADRA} predictions based on the measured  
distorted surface (dotted line). {\it Right:} Predictions for the cold
optics based on the values in Table~\ref{tab:surf_parms} for
Gaussian distortions (dashed line), hat box distortions 
(dotted line), and the undistorted mirror (solid line). At $\approx 2^{\circ}$ 
all curves meet.}  
\label{fig:srf_gain}
\end{figure*}

An approximation is made to compute the gain of the 
primary beam pattern for the full telescope.
First, $G_p$ is computed by substituting $l_c$, $\overline{\rho^2}$, 
and the ideal
$G_0$ into equation~\ref{E:Gauss_gain}.
Next $G_p$ is substituted for $G_0$ in Ruze's formula,
with $l_c$ and $\overline{\rho^2}$ for the secondary. The final result
is this newly determined $G(\theta)$. It is compared to the azimuthally 
averaged gain of the model. This method ignores the fact that
the primary is in the near field of the secondary.
Results are plotted for the warm reflectors in Figure \ref{fig:srf_gain}
along with measurements and predictions from the {\tt DADRA} code. 
For the warm reflectors, the simple model reproduces the overall 
shape but misses the lobe at $\theta=0.45^{\circ}$. 

\begin{figure*}
\epsscale{1.8}
\plotone{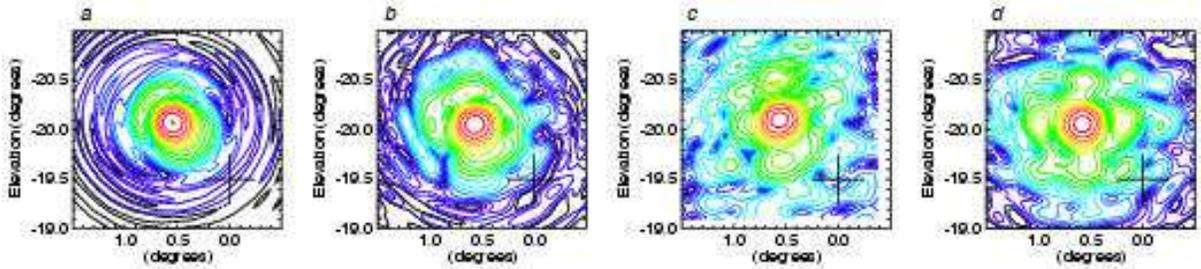}
\caption{A comparison of the W4A beams at 90 GHz. 
The panels show: (a) predictions for
an ideal-mirror beam pattern, (b) predictions for an ambient temperature 
distorted mirror beam pattern, (c) the measured beam pattern, and (d) 
the prediction for a cooled mirror in orbit. The contour intervals are 3~dB.
The cross indicates the center of the focal surface.}
\label{fig:4ws}
\end{figure*}

For the on-orbit predictions, the same procedure is followed except that
the cold $\sigma_z$ is used. We assume that the correlation length does not
change. The results are plotted in 
Figure \ref{fig:srf_gain}. The Ruze beam dominates the
undistorted beam for $0\fdg2 < \theta < 2^\circ$, and above $2^\circ$ is
subdominant to it, diminished by the exponential factor. 
Even $1\fdg4$ from the boresight, the Ruze beam is $<-35~$dB below
the peak. The full W-band patterns for the ideal and distorted beams
are plotted in Figure~\ref{fig:4ws}.   

An additional potential source of degradation of the main beams is print
through of the weave pattern that comprises the fabric that 
makes up the surface
of the reflectors. This was investigated by measuring the change in the
holographic pattern of a sample surface upon cooling \cite{mjackson}.
For a 200~K change, the contrast of the weave increases 
from 0 to $\approx 2~\mu$m.
From this pattern, one computes an upper bound of -70~dB in W-band for a narrow
sidelobe $40\deg$ off the main beam. Contributions from bright sources,
for example Jupiter, through this lobe will be $<0.1~\mu$K.  
 
\subsection{Sidelobes}
\label{sec:sidelobes}

	The sidelobes are small, well
characterized, theoretically understood, and measured. Generally, sidelobes
are produced by illumination of the edges of the optical elements
whereas the main beam profile is determined by the shape of the 
elements. The primary radiometric contamination comes
from hot sources, such as the Earth, Moon, and Sun (Table~\ref{tab:sem}),
well outside the main beam. Also, contamination arises from the Galaxy,
which is much colder, $\approx 200~$mK, but is extended and
closer to the main beam. For scale, if the Sun is to
contribute less than $2~\mu$K to any pixel, it must be rejected
at $(2~\mu{\rm K}/T_{Sun})(\Omega_{Sun}/\Omega_A)\approx
-100~$dB.  

	The six principal contributions to the sidelobes
are: a) response of the feeds to radiation outside the angle
subtended by the subreflector; b) diffraction from the edge of
the secondary; c) diffraction from the edge of the primary; d) spill past
the edge of the primary; e) scattering
from the structure that holds the feeds; and f) reflection off the
radiators by radiation that goes past the primary.
The more negative the edge taper, the smaller each of these is.

	The sidelobe levels are computed in two ways. For the
contributions from the radiators and optical surfaces, the full 
physical optics calculation is used \cite{YRS}. The commercial code 
was rewritten to take
into account the interaction of the radiators with the primary mirrors.
As the Sun, Earth, and Moon are shielded by the 5~m diameter solar 
shield, which is not part of the physical optics model, the geometric
theory of diffraction [GTD, Keller (1962)] is used to compute their 
contribution.
 
In GTD, the field diffracted by an edge is
given by $\psi_s(r) = D_{GTD}\psi_i (r(1+r/\rho_1))^{-1/2}e^{ikr}$,
where $\psi_i$ is the incident
field (or field at the edge), $r$ is the distance to the edge, $\rho_1$
is a radius of curvature that characterizes the edge\cite{Keller62}, and the 
diffraction coefficient $D_{GTD}$ is
\begin{equation}
D_{GTD}=-{e^{i\pi/4}\over2\sqrt{2\pi
k}}\biggl[{1\over\cos((\phi-\alpha)/2) } 
\pm {1\over\sin((\phi+\alpha)/2) }\biggr].
\label{eq:gtd1}
\end{equation}
For a straight edge $1/\rho_1=0$ and one recovers Sommerfeld's solution.
The angles are shown in Figure~6. 
The upper sign is for the incident E-field perpendicular to edge and the
lower sign is for the E-field is parallel to the edge.

Radiation from the Sun diffracts around the edge of the 
solar shield, 
diffracts over the top of the secondary, and then enters the feeds
as can be seen in Figure~\ref{fig:sideview}. Radiation from the Earth
and Moon diffracts directly over the top of the secondary and then enters
into the feeds. To compute these contributions, 
the rim of the solar shield is approximated as straight (outboard
of the secondary it is almost straight).
The edge of the secondary is approximated with a radius of 
curvature of $r_s = 40~$cm so that $1/\rho_1=-\cos(3\pi/2-\phi)/r_s$.
For all estimates, the phase in equation~\ref{eq:gtd1} is ignored.

Consider first the contribution from the Earth with temperature
$T_E$ and solid angle $\Omega_E$. The antenna temperature is given 
by
\begin{equation}
T_A = {|D_{GTD}|^2\over r } {G(\theta_{edge})\over 4\pi }T_E\Omega_E,
\end{equation}
where $G(\theta_{edge})$ is the gain of the feed at the edge of the
secondary, $r$ is the distance to the secondary, and $D_{GTD}$ is given by
equation~\ref{eq:gtd1} with the geometry in Figure~\ref{fig:sideview}.
The diffraction effects are greatest for K band because it has the
largest wavelength so only it is considered. Using the 
parameters in Table~\ref{tab:sem} and $G(\theta_{edge})=-22.4~$dBi,
$\alpha=-76^{\circ}$, $\phi= 200^{\circ}$, and $r=60~$cm, one finds
$D_{GTD}=0.36~$cm$^{1/2}$ and $T_A\approx 60~$nK. Similar calculations for
the Moon yield  $T_A\approx 4~$nK and for the Sun, after taking 
into account the double diffraction,  $T_A\approx 16~$nK.
\medskip
\centerline{\vbox{\epsfxsize=7cm\epsfbox{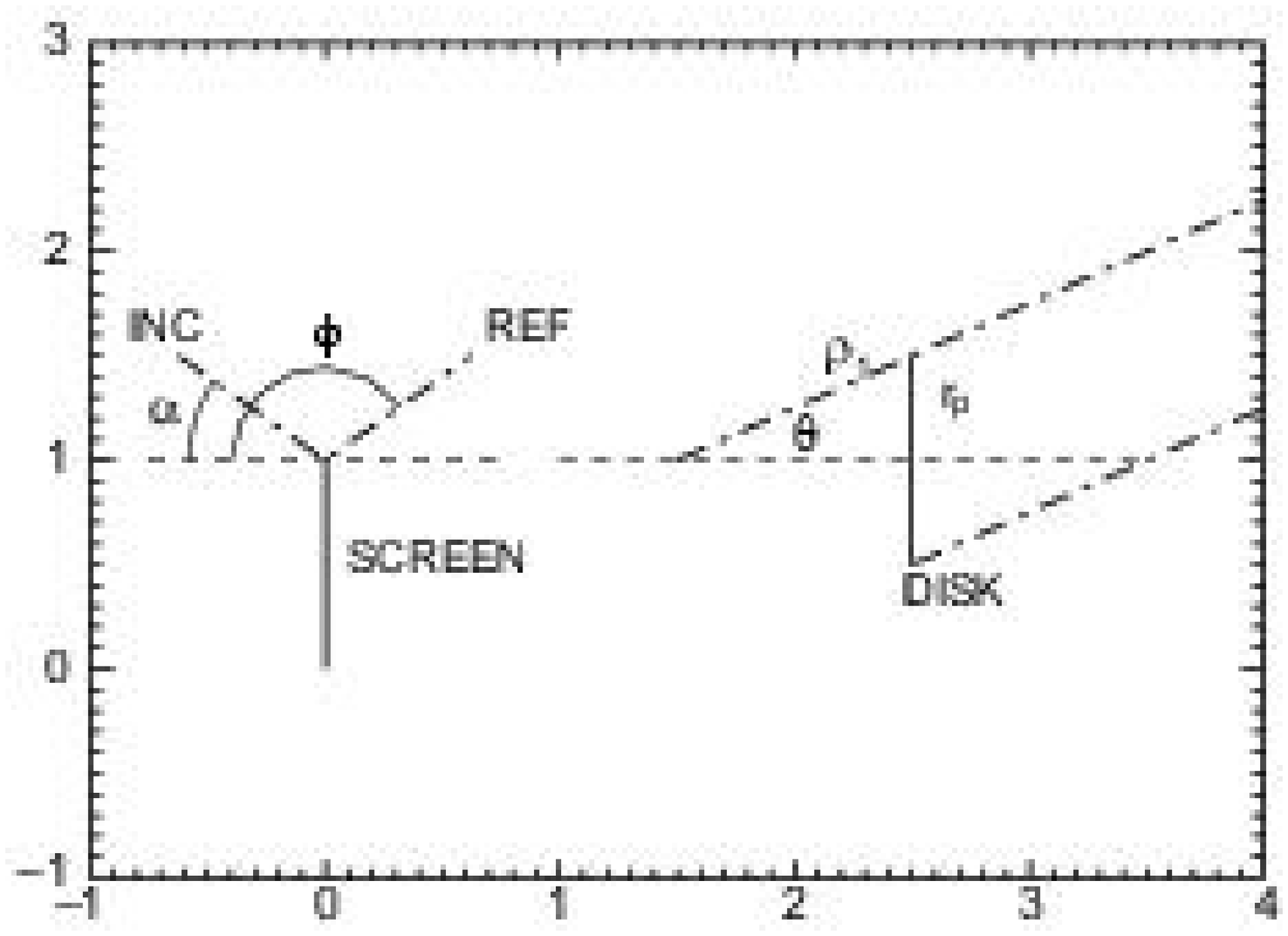}}}
\noindent{\small Fig. 6. 
The geometry for the diffraction calculations.
On the left, the incident and reflected rays are measured with respect 
to a line perpendicular to the screen (viewed edge on).
On the right is a disk, for example the primary. The dot-dashed 
lines are the two scattered rays from the edge.}
\medskip
\stepcounter{figure}

Similar order-of-magnitude calculations were used to check multiple
diffraction paths and angles. It was found that solar radiation could directly
enter the feeds by diffracting from the $\pm x$ edges of the secondary and
the truss structure that holds the secondary. As a result, additional
``diffraction shields'' were added that go between the front of the 
structure that holds the
feeds (FPA) and the edge of the secondary to eliminate these paths. 

GTD was also used early in the design phase in the parametric model 
of the satellite (Section~\ref{sec:intro}).
The illumination of the primary for the field $\psi$ was modeled 
as a Gaussian of width $\sqrt{2}\sigma_p$ with edge taper $y_e$ so that
\begin{equation}
\exp(-r_p^2/4\sigma_p^2) = 10^{y_e/20} <<1
\end{equation}
at the rim of the 
primary, $r=r_p$. For a circular disk or aperture evaluated in the 
limit of  $r>>r_p$ and intermediate angles
\begin{equation}
\psi_s(r,\theta) = D_{GTD}\psi_i (r^2\sin\theta / r_p)^{-1/2}e^{ikr},
\end{equation}
and
\begin{equation}
D_{GTD}=-{1\over2\sqrt{2\pi
k}}\biggl[{1\over\sin(\theta/2) } 
\pm {1\over\cos(\theta/2) }\biggr]
\end{equation}
At large angles, diffraction with E parallel to the edge dominates
and the full expression becomes
\begin{eqnarray}
\lefteqn{G(\theta) =  {4\pi r^2\psi_s^2(r,\theta)\over {\rm Total~power}}\approx}
\hspace{2.8truein} \nonumber \\ 
{\ln(10^{1/20})|y_e|10^{y_e/10}\over \pi r_p k}
{1\over\sin\theta}\biggl[{1\over \sin(\theta/2)} 
+ {1\over \cos(\theta/2)}\biggr]^2
\label{eq:gtd2} 
\end{eqnarray}
A curve of equation~\ref{eq:gtd2} is shown in Figure~\ref{fig:k_composite}.

\begin{figure*}
\epsscale{1.8}
\plotone{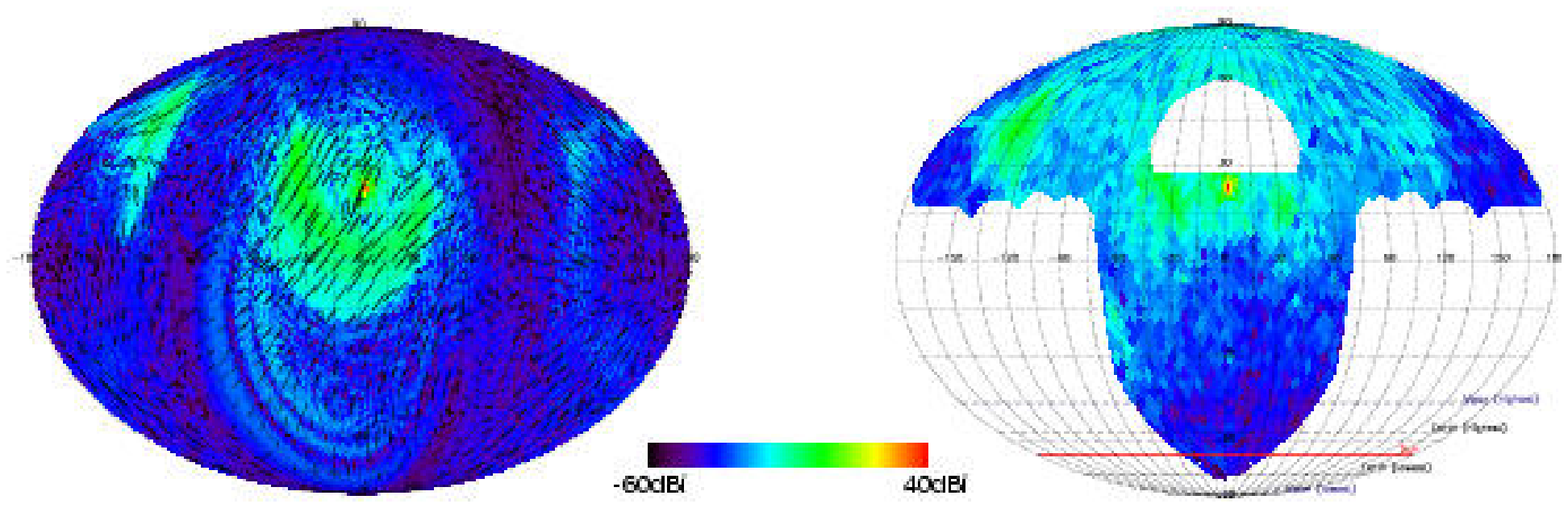}
\caption{{\it Left:} Predicted sidelobe response at 22 GHz over the full sky. 
The main beam is shown at an elevation of $17^{\circ}$. The forward
gain is 47~dBi and so is suppressed on this plot by almost $10^5$. The arrows
indicate the polarization direction. The large lobe in the upper left
side is from spill past the place where the primary and radiator 
meet on the left hand side of Figure~\ref{fig:curd}. The large rings are from
diffraction around the secondary. The lobe in the upper right is from
reflection off the radiator just below where the lobe on the left
passes through. Note that the most prominent lobe is polarized. Also,
K band has, by far, the largest sidelobes. This calculation does not
include any effects from the solar array panel. 
{\it Right:} Measurement of the sidelobes at the same frequency but with
the FPA and solar array in place. The hole in the upper central part 
is from incomplete 
coverage. The gross features agree. The extra power in the central upper
hemisphere, as compared to the prediction, is from
scattering off the top of the FPA. Because it is diffuse and at a low
level, $\approx -30~$dBi, it is not a significant source of pickup. The
solar array blocks all radiation in the lower hemisphere.}
\label{fig:Ksidelobes}
\end{figure*}

Because of the large uncertainty associated with the hand GTD
calculations and the need to confirm the computer models, 
the sidelobes were measured in K, Q, and W bands
using a full-scale replica of the satellite built around the REU. 
The results from the full diffraction calculation for K band
along with measurements at the same frequency 
are shown in Figure~\ref{fig:Ksidelobes}.  

\subsubsection{Radiometric Contributions From Sidelobes}

To assess the contribution to the 
radiometric signal from Galactic emission, the full sky differential beam
maps were ``flown'' over a galactic template.
The beam maps were made from a composite of the computer model and the 
measurements. The galactic template is based on the Haslam and IRAS 
maps scaled to be $T_{gal} = 230, 100, 70, 40, 50~$mK in K through W
bands respectively in a $4\times 10^{-6}$~sr pixel at $l=0, b=0$.

The A-side beam is placed in each pixel of the map and the negative
B-side beam is rotated
through $360^{\circ}$ in $10^{\circ}$ intervals. The contributions for
$G>15~$dBi are excluded as these are right next to the main beam 
and will naturally be incorporated into the map solution.
(For V and W band this corresponds to one pixel.)
At each orientation, the following
integration is performed:
\begin{equation}
T_A = {1\over 4\pi}\int G(\theta,\phi)T_{gal}(\theta,\phi)d\Omega
\end{equation}
where $G(\theta,\phi)$ is the telescope gain. For each ring, the minimum,
maximum, and {\rms} signals are recorded. Data with the B-side beam 
at $|b|<15^{\circ}$ are excluded. The results are given in 
Table~\ref{tab:sidelobes}.

Verifying the GTD calculations entails a measurement of the sidelobes at
the -70~dBi level. Measurements using the test range between the roof tops
of Princeton's physics and math buildings are limited by scattering at the 
-50~dBi level, 110 dB below the peak in W-band, even after covering
large areas of ground with microwave absorber. Consequently, only an upper 
limit of $0.5~\mu$K may be placed on contamination by the Sun, Earth,
and Moon. 

\begin{table*}[t]
\caption{Contributions and Parameterization of the Sidelobe response}
\small{ \vbox{
\tabskip 1em plus 2em minus .5em
\halign to \hsize {
     #\hfil & 
\hfil#\hfil &
\hfil#\hfil &
\hfil#\hfil &
\hfil#\hfil &
\hfil#\hfil \cr
\noalign{\hrule\vskip1pt\hrule\smallskip}
Band designation      &
   K            & 
   ${\rm K_a}$  & 
   Q        & 
   V        & 
   W\tablenotemark{a}  \cr
Center frequency (GHz)& 
   23    & 
   33    & 
   41    & 
   61    &
   93   \cr 
\noalign{\smallskip\hrule\smallskip}
Maximum ($\mu$K)  &
   $85$  & 
   $10$   & 
   $20$   & 
   $2$   &
   $1.5$ \cr
Minimum \tablenotemark{b}  ($\mu$K) &
   $-150$   & 
   $-60$   & 
   $-40$   & 
   $-3$   &
   $-5$    \cr
{\it rms} ($\mu$K)   &
   $12$   & 
   $2$   & 
   $3$   & 
   $0.2$   &
   $<0.1$  \cr
Max. Edge taper (dB)\tablenotemark{c}   &
   -13   & 
   -20  & 
   -21  & 
   -21   &
   -16/-20  \cr
Forward beam efficiency\tablenotemark{d} &
   0.960   & 
   0.986  & 
   0.986  & 
   0.996   &
   0.996/0.999 \cr
\noalign{\vskip1pt\hrule\vskip1pt\hrule\smallskip}
}}}
\tablenotetext{a}{Where appropriate, the values for W1/4 (upper) and
W2/3 (lower) are separately given.}
\tablenotetext{b}{The negative values correspond to a signal through 
the B beam.}
\tablenotetext{c}{The maximum edge taper for the center of the band.
As the current distributions are not symmetric with reflector, most of
the edge has a substantially smaller taper.}
\tablenotetext{d}{The integral of the beam in an area 
around $\pm 2^{\circ}$ from the maximum divided by $4\pi$. A value of 
0.996 means that 0.004 of the solid angle is scattered into the sidelobes.}
\label{tab:sidelobes}
\end{table*} 

\subsection{Reflector Surfaces}

	Coatings were applied to the optics to provide a surface 
a) whose microwave properties are essentially indistinguishable from
those of bulk aluminum, b) that radiates in the infrared so that
the reflector does not heat up unacceptably when the Sun strikes it,
c) that diffuses visible solar radiation 
so that sunlight is not focused on the focal plane or the secondary, 
and d) that does not allow the build up of excessive
amounts of charge in orbit. These four criteria can be met 
simultaneously \cite{Heaney}.

\begin{figure*}
\epsscale{1.6}
\plotone{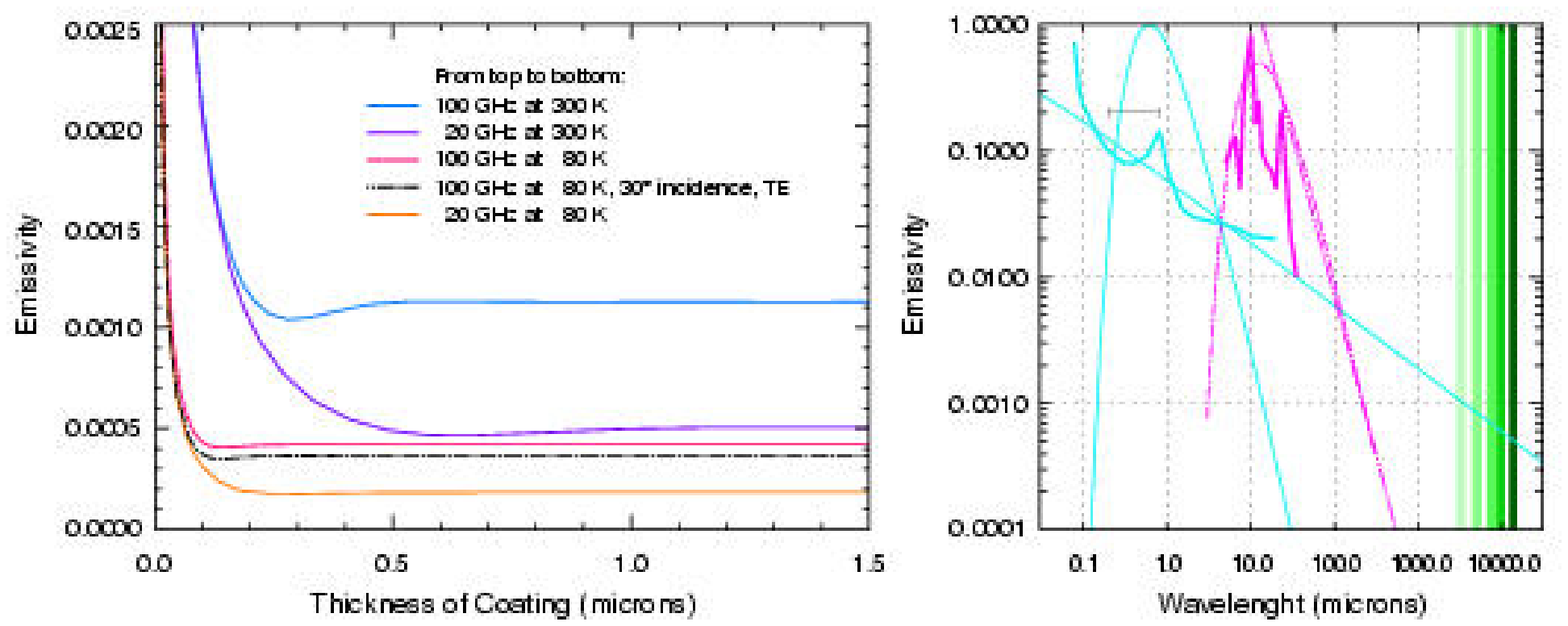}
\caption{{\it Left:} The theoretical emissivity of a thin aluminum
coating as a function of the thickness of the coating. The dip in
the curve is an interference phenomenon (Born \& Wolf pg. 628). 
 At small coating thicknesses, the results differ from
those in Xu {\it et al.} (1996) and give a somewhat smaller thickness at the
minimum than
Garg {\it et al.} (1975) who find $t/\delta\approx \pi/2$, where $\delta$ is
the skin depth. The variation in emissivity with 
incident angle is
not significant. The dot-dot-dot-dash line is for the TE mode
(E perpendicular to plane of incidence) with a $30^{\circ}$
incident angle. The TM mode will be higher than the nominal
value.  The effects of the composite substrate are negligible
(e.g., Ramo \& Whinnery pg 249).
{\it Right:} The emissivity of the surface from optical to microwave
wavelengths. The aluminum emissivity is based on the Drude
model (solid straight line, {\it e.g.}, Ashcroft \& Mermin (1976)) 
and measurements (curve with
a peak at $1~\mu$m). The visible band is indicated by a bar at an
emissivity of $e=0.2$,
just below the peak of a 6000~K blackbody representing the Sun. The
jagged line at 10~$\mu$m is the emissivity of ${\rm SiO_2}$ and the 
dot-dot-dot-dash line is a model of its long wavelength behavior. The
vertical bands on the right are the {\sl MAP} bands}
\label{fig:refl_emis}
\end{figure*} 

	The coating is vacuum deposited onto a
surface comprised of an $\approx 25~\mu$m thick
layer of epoxy over a 250~$\mu$m composite sheet. 
The epoxy layer is roughened to diffuse sunlight and not 
affect the microwave properties,
and is then coated with aluminum and SiO$_x$ (the  ``$x$'' denotes unknown,
as the material is a combination of SiO and SiO$_2$). The SiO$_x$ coating,
because of an absorptance resonance,
allows the reflector to radiate at $10~\mu$m, near the peak of
a 300 K blackbody, as shown in Figure~\ref{fig:refl_emis}.
To maximize the infrared radiation, thicker
SiO$_x$ is better, but if the layer exceeds $2.5~\mu$m it insulates the
aluminum unacceptably leading to potential problems with surface charging
in space.

	Obtaining a successful coating was particularly challenging
\footnote{For example, the REU reflectors turned brown
over a period of days shortly after delivery. Later, a coating 
that was apparently
stable over a year peeled off the flight secondaries and had to be redone.}.
The mission requirement 
that the optics be able to withstand transient direct illumination by the Sun
was a significant complication. To maintain high surface quality,
especially in the micron wavelength region, the flight reflectors 
were maintained at $<50$\% relative humidity from when they 
were coated until launch.
 
\subsubsection{Roughening the Surface}

The reflector surfaces are roughened to diffuse the solar
radiation so that the secondaries and feeds do not get too hot
from the focused radiation from the primary. 
The specification, which is developed below, is that no more than
20\% of the reflected radiation be inside the $20^{\circ}$ (full angle)
cone of the reflected ray at a wavelength of 540 nm. 
Measurements show this is possible
if the surface has a $0.4~\mu$m roughness with a correlation length
of $30~\mu$m, in agreement with the models \cite{Bennett61}.

Reflection from a roughened surface is quantified with 
the bidirectional reflectance (BDR), $\rho^\prime$, 
and $\rho_d$, the directional reflectance (DR)
\cite{Nicodemus,BeckSpizz,Davies,HH67}. The BDR is the reflection coefficient
per unit solid angle for arbitrary incident polar directions ($\psi,\zeta$) and
reflection  directions ($\theta$,$\phi$) measured from the mean normal
of the surface:
\begin{equation}
\rho^\prime(\psi ,\zeta ;\theta ,\phi ) \equiv {\delta I_r(\theta,\phi ) 
\over I_i(\psi,\zeta ) \cos\psi \Delta\omega_i} = 
{\pi ({\rm Energy~into~} \Delta\omega_r)  \over {\rm Total~incident~energy}},
\end{equation}
where $I$ is the spectral intensity in W\,sr$^{-1}$Hz$^{-1}$ 
and $\Delta\omega_i$
and $\Delta\omega_r$ are the solid angles containing the incident and
reflected radiation. The units are sr$^{-1}$. In practice, the BDR is not
measured absolutely and $\zeta$ and $\phi$ are $180^{\circ}$ apart.
In terms of the parameters for a lossless surface:
\begin{eqnarray}
\lefteqn{ \rho^\prime(\psi ,\zeta ;\theta ,\phi) = 
{\exp(-F) \over\cos(\psi ) \Delta\omega_i}+ }\hspace{3.5truein}  \\
{\exp(-G)\over\cos(\psi )\cos(\theta_i )}\pi 
\biggl( {c\over\lambda}\biggr)^2\,B
\sum_{m=1}^{\infty} {G^m\over m\cdot m!}
\exp\biggl[-{\pi^2\over m }\biggl({c\over \lambda }\biggr)^2H\biggr] 
\nonumber \\ 
{\rm where~~} F = (4\pi{\sigma\over\lambda} \cos \psi)^2 
~~\& ~~G = (2\pi{\sigma\over\lambda} (\cos\psi + \cos\theta))^2. 
\hspace{0.05in}\nonumber
\label{eq:bdr}
\end{eqnarray}
Here, $\sigma$ is the surface {\it rms}, $c$ is the correlation length,
$\Delta\omega_i$ is the solid angle for the incident radiation, $B$
is a function of the incident angles and is of order one, and
$H =[\sin^2\psi + \sin^2\theta + 2\sin\psi\sin\theta \cos(\zeta-\phi)]$.
When there is no scattering, the first term dominates and the reflection
is specular and coherent. The second term gives the incoherent 
or diffuse component.
The correlation length enters the incoherent component alone and so
influences only the spatial distribution of the
radiation and not the total energy reflected \cite{HH67}, similar to the case of
Ruze scattering. One expects that for a fixed surface {\it rms}, the greater
the coherence length, the smaller the slope, and the more confined the
reflected beam. For a perfectly diffuse
(Lambertian) reflector, $\rho^\prime$ is independent of angle.

The DR is the ratio of the total energy reflected
into a hemisphere divided by the incident energy,
\begin{equation}
\rho_d = {\int\int I_r(\theta,\phi)sin(\theta)cos(\theta)d\theta
d\phi \over I_i(\psi )sin(\psi)cos(\psi)d\psi d\zeta},
\end{equation}
and is typically 0.6. In practice, the DR is the measured quantity and the BDR
is estimated from it through $\rho^{\prime}=\rho_d/\pi$. 
Thus, for a lossless Lambertian reflector, $\rho^\prime=1/\pi$ in a plot
of the BDR. For the {\sl MAP} reflectors, the reflection is d
iffuse and incoherent for $\lambda < 2~\mu$m and thus the 
specular component may be ignored.
An example is shown in Figure~\ref{fig:bdr}.
The incoherent scattering falls as $(c/\lambda)^2\approx 10^{-4}$  
at microwave wavelengths and so it is negligible and has no 
effect on the beams. 

\subsubsection{The SiO$_x$ Coating and Optics Temperatures }

Because of its relatively good absorptance at 1 $\mu$m
(Figure~\ref{fig:refl_emis}), near the peak
of the solar spectrum, aluminum gets hot in the Sun. 
For a flat plate normal 
to the Sun that emits only on the illuminated face, 
the radiative steady state temperature is
\begin{equation}
 T = \biggl({\alpha_{sol}F_{sol}\over \epsilon_{hemi} \sigma_B}\biggr)^{1/4} 
= 398 \bigg({ \alpha_{sol}\over \epsilon_{hemi}}\bigg)^{1/4}~{\rm K} 
\approx 580~{\rm K}
\end{equation}
where $\sigma_B = 5.67\times10^{-8}$~${\rm W m^{-2}K^{-4}}$,
$\alpha_{sol}\approx0.085$ is the solar absorptance for optical quality
aluminum at room temperature, $\epsilon_{hemi}\approx  0.018$ is 
the total hemispherical
emittance (coefficient for the total emitted flux into
$2\pi$~sr) for aluminum at room temperature, and the 
maximum solar flux (at perihelion) is 
$F_{sol}=1420~{\rm Wm^{-2}}$ near the Earth. This temperature is well
above the glass
transition temperature of the FM73 epoxy in the reflectors, 
370~K. To cool the reflectors, they are coated with $\approx 2~\mu$m
of SiO$_x$, which emits strongly near 10~$\mu$m as shown in 
Figure~\ref{fig:refl_emis} \cite{Hass}. This results in  
$\alpha/\epsilon\approx 0.8$ and so $T=376~$K for extended
direct illumination. Because the SiO$_x$ is so thin, it has no 
effect on the microwave properties of the reflectors \cite{MonSievers75}.

The secondaries see focused radiation from the Sun,
which would increase the flux on them to $\approx 10F_{sol}$ were the
primaries perfect reflectors. However, the sunlight is diffused
and absorbed by the primaries which substantially decreases the flux on the
secondaries. To ensure that the secondaries do not exceed the glass transition,
the flight optics were illuminated with a stage spot light and the flux at
the position of the secondary was measured. After correcting for the
geometry and spectral difference between the test light and the Sun, 
the data were 
compared to a computer model of the scattering that also predicted the
Sun's net effect. The computer output was then input to a full thermal
model of the secondary and S/C that accounted for the rotation of the S/C 
during solar exposure. The worst case transient 
temperature predictions were within 10~K
of the glass transition temperature. This margin
was considered sufficient. As a guideline to the full model, the
temperature of the secondary is approximately given by
\begin{eqnarray}
T_{sec} = \biggl({\alpha^{sub}_{sol} \over \epsilon^{sub}_{hemi}
}\biggr)^{1/4} \biggl({F_{inc}\over\sigma_B}\biggr)^{1/4}
= \biggl(f_{i20} {\alpha^{sub}_{sol} \over \epsilon^{sub}_{hemi}
}\biggr)^{1/4} 507~{\rm K}\hspace{0.5truein} \nonumber \\
{\rm with}~~~F_{inc} = \biggl({140~{\rm cm}\over 45~{\rm cm}}\biggr)^2
(1-\alpha^{main}_{sol})F_{sol}
\biggl( {13.5^{\circ} \over 20^{\circ}}\biggr)^2 f_{i20}
\hspace{0.5truein}
\end{eqnarray}  
where $(140/45)^2$ is the ratio of the spot size on the primary to that
on the secondary, $\alpha_{sol}^{main}\approx 0.4$,
$(13.5/20)^2$ is the ratio of the angle of the solar
spot on the secondary as viewed from the primary to the $20^{\circ}$
reference angle, and $f_{i20}$ is the flux within $20^{\circ}$ as
quantified with equation~\ref{eq:bdr}. The specified values for the
secondaries were $f_{i20}<0.2$ and $\alpha/\epsilon<0.9$ resulting
in $T_{sec}\approx 330~$K and a flux of $\approx F_{sol}/2$.
These values are close to those of the full model.
A spare secondary was tested with a solar simulation and was found
to be able to survive temperatures of $115\deg$C for periods of
$\sim 2~$min.

\begin{figure*}
\epsscale{1.5}
\plotone{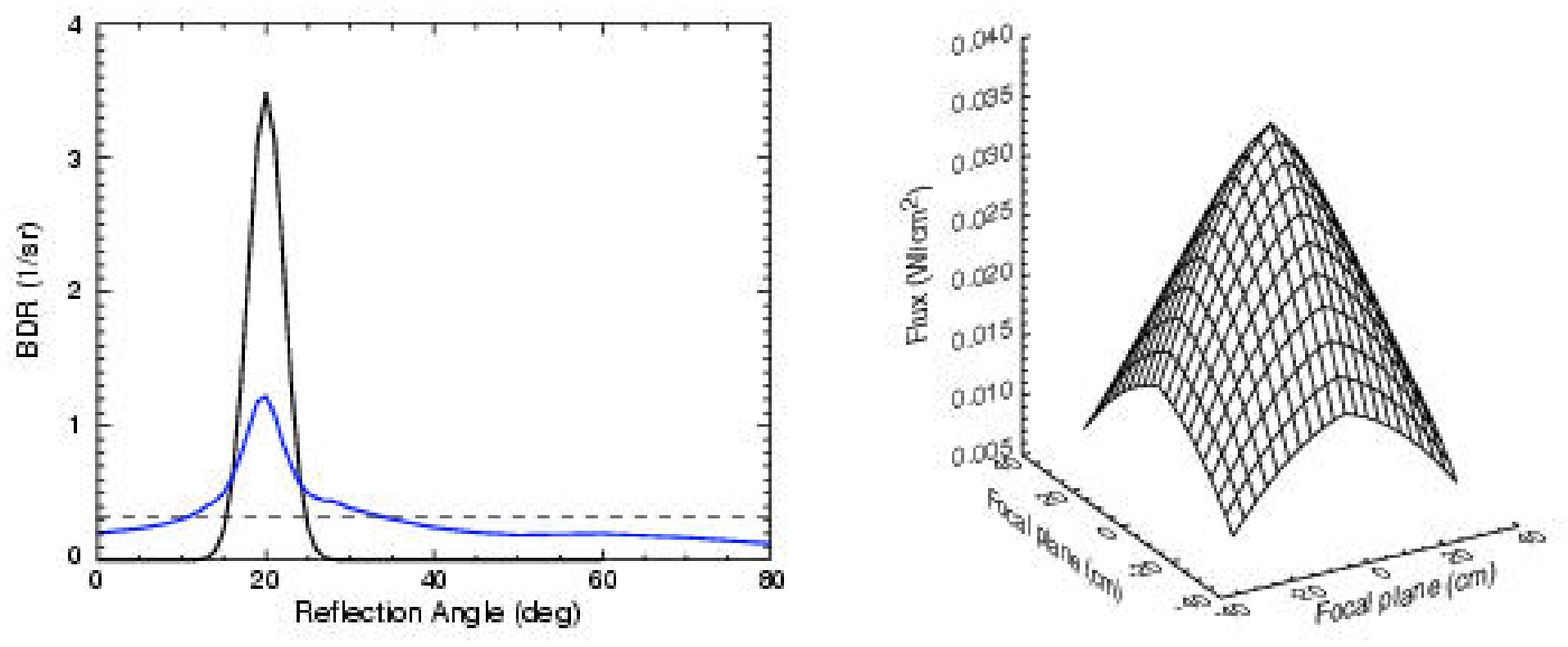}
\caption{{\it Left:} The model incoherent BDR (higher peak) and 
the measured BDR for primary P3. The model is from 
equation~\ref{eq:bdr}, with $\sigma = 0.4~\mu$m, $c= 30~\mu$m, 
$\lambda= 0.5~\mu$m,
and $\psi = 20^{\circ}$, scaled by 0.09 to match a
typical DR. This approximation misses the wings but
gets the peak sufficiently well. For the surface upon which the model is
based, 13\% of the reflected energy is inside
a $20^{\circ}$ cone; for P3 this quantity is 11\%.
The dashed line at $1/\pi$ is for a lossless Lambertian surface.
{\it Right:} Distribution of power in the 
focal plane after reflecting off both reflectors.}
\label{fig:bdr}
\end{figure*}

The diameter of the solar image in the focal plane is 2.2 cm. 
With perfect reflectors, the flux would be $\approx 2600~F_{sol}$
and would vaporize the aluminum feeds.
To estimate the actual flux, equation~\ref{eq:bdr}
is convolved with itself after accounting for the geometry
and absorptance of the reflectors. The result is shown in 
Figure~\ref{fig:bdr}. With a maximum flux of $\approx F_{sol}/4$, 
a simple model of the conductance and emittance of the feeds shows 
that they should not exceed $40^{\circ}~$C in space.
Measurements with an intense spotlight were consistent 
with the calculations. 

\subsubsection{Microwave Properties of the Aluminum Coating}

Aluminum was chosen for the metallic reflector coating
because its deposition is well studied and its
microwave emissivity is sufficiently low for {\sl MAP}.
For all emissivity calculations, the classical regime obtains:
there are no anomalous effects \cite{Pippard} and the displacement 
current can be ignored. Specifically, 
$\epsilon = 4\pi\delta/ \lambda  = ( 16\pi\epsilon_0f/ \sigma)^{1/2}$.
Thus the emissivity scales as
$f^{1/2}$. At 70~K the emissivity at 100 GHz is about 0.0004
($\sigma_{300~{\rm K}}/\sigma_{70~{\rm K}}= 0.15$).
For two surfaces (primary and secondary) at 70 K, the contribution
to the system temperature from the thermal emission is 64 mK at
100 GHz, which is negligible.

Butler (1998) made a precise room temperature measurement of the 
difference in emissivity between machined stainless steel, copper, aluminum,
and vacuum deposited aluminum (VDA) sample with an SiO$_x$ coating. 
The sample has the same construction as the {\sl MAP} reflectors
except that it is flat. Butler
found the differences in emissivity between all materials and the 
VDA sample to be within
20\% of the value computed assuming a conductance of bulk aluminum 
for the sample. In other words, the VDA coating acts like bulk aluminum
at microwave frequencies. Numerous coating
samples were checked. No degradation in the microwave properties was
observed over a period of a year.
No change was seen after thermal cycling to 77~K.

To ensure the similarity of the reflectors, the coating procedure specified
that the same prescription be used for symmetric pairs of reflectors.
From Figure~\ref{fig:refl_emis}, 
one sees that with $t>1.2~\mu$m, the bulk emissivity should be
obtained. To be insensitive to any small variations in the thickness, 
small defects in the composite surface, or interactions with the
composite substrate, a thickness of $t>2.2~\mu$m
was specified. This is many times the 90~GHz skin depth of $0.11~\mu$m
at 70~K.

	The calculation for Figure~\ref{fig:refl_emis} is 
based on the emissivity of
a single thin sheet of aluminum at normal incidence. One can be more ambitious
and include the XN70 substrate but the results are essentially 
unchanged: the net
emissivity is dominated by the outer most layer of aluminum.
The dominant effect in determining the emissivity is the impedance
mismatch between the vacuum and the aluminum, for which the index
$n_{Al}\approx 2000+i2000$ at 90 GHz.  

\subsubsection{Spacecraft Charging}

	Surface charging is a well-known and potentially serious
problem for spacecraft. In short, 
current from the ambient plasma can charge the external surfaces
of the S/C to very high potentials ($>10~$kV) relative to space
or to other spacecraft surfaces. These charged surfaces 
are subject to abrupt discharge either to a spacecraft surface
at a different potential or to space itself. Such high-potential
discharges can wreak havoc on the electronics or damage sensitive 
surfaces. The problem is complicated and
empirical. The physical processes include photo ionization, 
electron and proton current
densities, secondary electron densities produced by collisions with the S/C,
space charge regions, etc. \cite{Jursa85}.  

The surface charging is large when the spacecraft is
between 6-10 Earth radii (geosynchronous orbit is 5.6 $R_E$) and
when there is high geomagnetic activity. Current flow from 
photoionization by the 
Sun tends to lower the magnitude of charging from the ambient plasma.
Thus, a classic scenario for a discharge event would be entry into 
(or emergence from) eclipse into sunlight.
It is believed to be 
possible to reach $-25,000~$V (relative to infinity), though $-5000~$V is more
typical. The plasma energies are typically about 10 keV and the currents
into the spacecraft are of order 1 nA/cm$^2$.  
The L$_2$ environment ($\approx 240~R_E$) is affected by the
interaction of the Earth's magnetic field and the solar wind
\cite{Evans02}.
As the orientation between the Earth and Sun changes, L$_2$ moves
through different charging environments the most dangerous of which is
believed to be the magnetosheath. It is possible that {\sl MAP} will experience
$\leq 25$~keV electrons with a current density of 
0.1 nA/cm$^2$ from this source.

If the SiO$_x$ is too thick, it insulates the primary. 
The NASCAP program \cite{Jursa85} indicates that a $2.5~\mu$m layer charges
to $\approx -160~$V. Tests
made on a 150~K sample surface with charged contacts and an electron
beam showed that the surface would not abruptly arc with up to 450~V and
currents $\approx 10$~nA/cm$^2$. The surface instead discharges in a
self limiting manner.
\section{Performance and Characterization of Optical Design}

{\sl MAP}'s radiometric sensitivity is a result of intrinsically low-noise
transistors in a HEMT amplifier that operates over roughly a
20\% bandwidth \cite{Posp00,Jarosik03}. The characterization of the optical system takes
this bandwidth into consideration. In radio astronomy parlance, {\sl
MAP} is a ``continuum'' receiver.
{\sl MAP} is characterized primarily in flight through measurements
of the CMB dipole, planets, and radio sources. 
Because of the wide bandwidth, sources with different
spectra have different effective frequencies and beam sizes, even within
one band.

Two fundamental assumptions about celestial sources are made that 
allow one to characterize the response as an integral of
the response at each frequency. They are:
a) signals received at different frequencies are
incoherent and b) signals received from different points on a source 
are incoherent \cite{TMS}.

In a perfectly balanced differencing assembly (DA), 
the output of any one detector switches at 2.5 kHz
from being proportional to the power entering side A
to the power entering side B. The A$-$B 
response is synchronously demodulated, integrated for 25.6 ms, 
and averaged. With the 2.5 kHz phase switch in one position, the power
delivered to a diode detector, from one polarization of one feed
when viewing a source of surface brightness $S_{\nu}$, is given
by
\begin{eqnarray}
W(\alpha,\beta,\gamma) = {1\over 2}\int
\int\alpha_l(\nu) \eta_{R}(\nu )A_e(\nu )f(\nu)\,
S_{\nu}(\vec\Omega)\,  \nonumber \\ 
\times B_{n}(\nu ,R_{\alpha\beta\gamma}\vec\Omega)\, 
d\Omega d\nu ,
\label{eq:www}
\end{eqnarray}
where $f(\nu )$ is the normalized bandpass of a DA at a
reference plane at the OMT/feed interface,
and $\alpha_l$ accounts for the loss in the microwave components. 
The feed and antenna losses are included in the radiation efficiency 
$\eta_{R}$. The normalized beam power
pattern $B_n(\nu,\vec\Omega)$, and the effective
collecting area of the antenna at normal incidence, $A_e(\nu )$, 
have been expressed separately. Where appropriate in the following, 
their product is written as $A_{eB}(\nu,\Omega )$.
$S_\nu(\vec\Omega)$ is a surface brightness with 
units of ${\rm W\,m^{-2}sr^{-1}Hz^{-1}}$
and is defined with respect to a fixed coordinate system. 
The beam is measured in its own coordinates and
$R_{\alpha\beta\gamma}$ is the matrix that specifies the beam position 
and orientation on the sky.

The gain of the optics is measured with a standard
gain horn at the GSFC/GEMAC facility. The measurement accuracy
is $\pm 0.2$~dBi. The peak gain is measured at 500 frequencies across
the band. Full beam maps are
made at twelve frequencies per band. The outer two
measurements are approximately 10\% beyond the nominal passband,
two more are at the band edges, and the remaining eight are
equally distributed across the band. No phase information is
required or used in the analysis.

The loss in the system comes from the optics, $\eta_{R}$,
and the radiometer chain, $\alpha_l$. In practice, the overall level of
both of these is calibrated out and so only their frequency dependence affects
the characterization. The dependence of $\alpha_l$ is accounted for 
in the measurement of $f(\nu)$ and so it is dropped in the following.
The dependence on $\eta_{R}(\nu)$, which is expected to be small, is found by
calculation and so it is retained. For the values in the tables, we take
$\eta_{R}=1$ across the passband.   

\subsection{Response to Broadband Sources}

There are a number of possible definitions of the effective area,
gain, frequency, and bandwidth. We choose ones that 
reduce naturally to those for a thermal emitter.
To this end, a source is modeled as 
\begin{equation}
S_{\nu}(\vec\Omega) = \sigma(\nu)S(\vec\Omega),
\end{equation}
where $\sigma(\nu)$ is 
$\propto \nu^{-0.7}$ for synchrotron emission, 
$\propto \nu^{-0.1}$ for bremsstrahlung or free-free emission, 
$\propto \nu^{2}$ for Rayleigh-Jeans emission, 
$\propto \nu^{3.5}$ for dust emission, and a Planck blackbody at
2.725~K\footnote{ The effective temperature scaling is obtained from 
$\sigma(\nu)$ by subtracting 2 from the exponent.}. 
The total power received is 
\begin{equation}
W(\alpha,\beta,\gamma) = {1\over
2}\int A_e^{bb}(R_{\alpha\beta\gamma}\vec\Omega)
S^{bb}(\vec\Omega) d\Omega,
\end{equation}
where the broadband effective area is defined as
\begin{equation}
A_{e}^{bb}(\Omega) \equiv { \int \eta_R(\nu)f(\nu)\sigma(\nu)A_{eB}(\nu,
\Omega)d\nu \over  \int \eta_R(\nu) f(\nu )\sigma(\nu)d\nu }.
\label{eq:aeff}
\end{equation}
and the frequency weighted source function is defined as
\begin{equation}
S^{bb}(\vec\Omega ) 
\equiv S(\vec\Omega)\int \eta_R(\nu) f(\nu) \sigma(\nu) 
d\nu ~~~{\rm (W/m^2\,sr)}.
\end{equation}
The effective area is never measured and is only defined through 
equation~\ref{eq:aeff}. If the source is a spatially uniform 
Rayleigh-Jeans emitter at temperature $T$, then $\sigma(\nu) = 2\nu^2
k_B/c^2$ and with a flat passband
\begin{equation}
W=(k_B\,T/c^2)\int\int\nu^2 A_{eB}(\nu,\vec\Omega)d\Omega 
d\nu  = k_BT\Delta\nu.
\end{equation}

The broadband gain is defined as
\begin{equation}
g^{bb}(\vec\Omega) \equiv {4\pi A_e^{bb}(\vec\Omega) \over 
\int A_e^{bb}(\vec\Omega^\prime)d\Omega^\prime }
=
{\int \eta_R(\nu)f(\nu)\sigma(\nu)g(\nu,\vec\Omega)/\nu^2\,d\nu 
\over \int \eta_R(\nu)f(\nu)\sigma(\nu)/\nu^2\,d\nu  }
\label{eq:gbb}
\end{equation}
where $g(\nu,\vec\Omega)$ is the quantity given by computer models
of a lossless system. Any measurement includes an error in the 
calibration, $\pm0.2$~dBi, and antenna loss $\eta_R(\nu)$,
though the reflector loss is of marginal importance even in W band.
Generally, it is found that the measured forward gain  $g_{m}^{bb}$ 
is $\approx 0.3$~dBi lower than ideal gain, $4\pi/\Omega_m^{bb}$, 
as a result of some loss and slightly more scattering out of the main 
beam than is accounted for in the models.

The effective frequency of the radiation is given by
\begin{equation}
\nu_e \equiv  
{ \int \nu f(\nu) 
\eta_{R}(\nu)A_{e}(\nu)\sigma(\nu) d\nu \over
\int f(\nu) \eta_{R}(\nu)A_e(\nu)\sigma(\nu) d\nu }.
\label{eq:nubb}
\end{equation}
Implicit in this definition,
and others in which $A_e$ explicitly appears, is that the source is
smaller than the beam. If a source fills $4\pi~$sr, then there would be
no dependence on the beam solid angle. This is the reason that
the numbers reported here are different from those in Jarosik {\it et
al.} (2003). The effect is that as one probes higher $l$ the effective center
frequency, and therefore the Rayleigh-Jeans to thermodynamic 
correction, changes.

\begin{figure*}
\epsscale{1.8}
\plotone{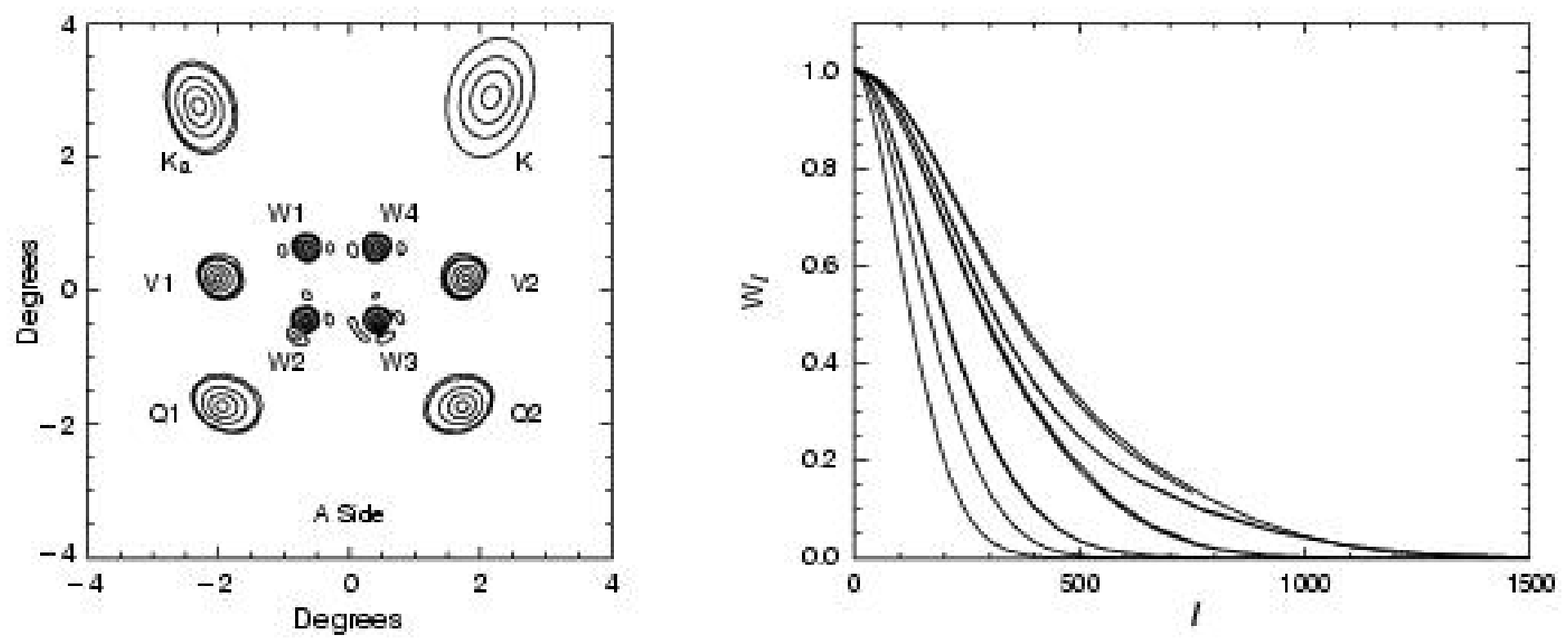}
\caption{{\it Left:} The predicted beam profiles for cooled and distorted optics
for a thermal source. If the beams could illuminate the sky, 
this is the pattern one
would see projected on the sky as viewed from the spacecraft. If one
looks at the B side focal plane array, as shown in
Bennett {\it et al.} (2003), the corresponding feeds are left-right
reversed. The contours are 0.9, 0.6, 0.3, 0.09, etc. the beam maximum. 
The small lobes off the 
W-band beams are the result of the reflector distortions which are
greater on the B side than the A side. The lowest contour in W-band 
is at 0.03. With the 
measurements combined with the computer code, the beams may be 
modeled to sub-percent accuracy. 
The beams in flight will be additionally smeared due
to rotation of the satellite. {\it Right:} The predicted B-side window functions
adjusted for rotational smearing. From left to right on the
plot, the windows correspond to K through W ``lower'' and then 
the W ``upper'' beams.}
\label{fig:beam_maps}
\end{figure*}

For estimating antenna temperatures and power levels, 
the effective bandwidth is convenient. It is defined as
\begin{equation}
\Delta\nu_e \equiv  \biggl( 12 {\int f(\nu)\eta_{R}(\nu) A_e(\nu)\sigma(\nu) 
(\nu-\nu_e)^2 d\nu \over \int f(\nu)\eta_{R}(\nu) A_e(\nu)\sigma(\nu)
d\nu}\biggr)^{1/2}  
\label{eq:dnubb}
\end{equation}
The factor of 12 makes a flat bandpass have a width of
$\Delta\nu$. This quantity does not
enter into any calculation and is distinct from 
the noise passband \cite{Dicke46}, $\Delta_n\nu$, quoted 
in Jarosik {\it et al.} (2003).

The flux from point-like radio sources is given as  
$F_\nu = \int_{\Omega}S_\nu(\Omega) d\Omega$ where
$F$ is measured in Janskys ($10^{-26}$ W\,m$^{-2}$Hz$^{-1}$).
For a narrow frequency band,
\begin{equation}
T_A\Delta\nu = {1\over 2k_{B}}\int {G_m(\nu)c^2\over
4\pi\nu^2}F_\nu d\nu
\rightarrow {A_e\over 2k_{B}}F_\nu\Delta\nu
\end{equation}
The usual notation is  $T_A = \Gamma F_{\nu}$. In the literature,
the flux is modeled as $F_\nu \propto (\nu/\nu_e)^\alpha$ where
$\alpha=2$ for a Rayleigh-Jeans emitter and
$\nu_e$ is given by equation~\ref{eq:nubb}.
Since $\Gamma$ depends on
the illumination of the primary, it will be similar for all bands. For
a ``flat spectrum'' source, $F_\nu$ is independent of
frequency ( $\alpha=0$) and has
approximately the signature of free-free emission. Such a source has the
same approximate antenna temperature in each band.

Implicit in the above is that $B_n = 1$ at all frequencies.
In other words, there is no dependence on the beam. The broad band 
conversion factor is given by
\begin{equation}
\Gamma^{bb} = { (c^2 / 8\pi k_B\nu_e^\alpha )
\int f(\nu)\eta_R(\nu) G_m(\nu)\nu^{\alpha-2}d\nu  \over 
\int f(\nu) d\nu}
\label{eq:jansk}
\end{equation} 

From the twelve beam measurements the beam characteristics for each
source are computed. For the 
CMB, the predictions of $g^{bb}$ are shown in Figure~\ref{fig:beam_maps} for
what is expected at L$_2$. Table~\ref{tab:measbeams} shows how the
effective broad band quantities depend on the source. The actual flight
values will be different. The ellipsoidal 
shape of the low frequency bands results from their large distance from 
the optimal focal position. The {\sl MAP} scan strategy has the effect 
of symmetrizing the beams, mitigating some of the effects of asymmetric beams. 
The small lobes in the cross pattern on the W and V
band beams are a result of the deformations in the mirror from the cool
down distortion. Though the distorted beams will complicate the analysis, 
they are well understood, can be modeled to the sub percent level, and will be
measured multiple times in flight. 

\begin{table*}[t]
\caption{Typical Pre-flight Effective Center Frequencies, Passbands, \& Gains
for Ideal Beams}
\small{ \vbox{
\tabskip 1em plus 2em minus .5em
\halign to \hsize {
     #\hfil & 
\hfil#\hfil &
\hfil#\hfil &
\hfil#\hfil &
\hfil#\hfil &
\hfil#\hfil &
\hfil#\hfil \cr
\noalign{\hrule\vskip1pt\hrule\smallskip}
Band designation \tablenotemark{a} &
   K            & 
   ${\rm K_a}$  & 
   Q            & 
   V            & 
   W$_{low}$    & 
   W$_{up}$     \cr
\noalign{\smallskip\hrule\smallskip}
Freq range (GHz) & 
   20-25    & 
   28-37    & 
   38-46    & 
   53-69    &
   82-106   &
   82-106   \cr 
Noise bandwidth, $\Delta_n\nu$ (GHz) & 
   5.2    & 
   6.9    & 
   8.1    & 
   10.5    & 
   19.0    &
   16.5   \cr 
 \%Band  & 
   22.6    & 
   20.6   & 
   19.6    & 
   17.2    & 
   20.2    &
   17.6   \cr 
\noalign{\smallskip\hrule\smallskip}
Synchrotron: $S_{\nu}\propto \nu^{-0.7}$   \cr
$\nu_e$  (GHz) &
   22.4  & 
   32.6   & 
   40.5   & 
   60.5   &
   94.9   &
   93.0 \cr
$\Delta\nu_e$ (GHz) &
   5.7  & 
   7.2   & 
   8.5   & 
  13.9   &
  22.3   &
  17.7   \cr
\noalign{\smallskip\hrule\smallskip}
Free-free: $S_{\nu}\propto \nu^{-0.1}$   \cr
$\nu_e$  (GHz) &
 22.5  & 
 32.7   & 
 40.6   & 
 60.7  &
 93.0  &
 93.1   \cr
$\Delta\nu_e$ (GHz) &
 5.7  & 
 7.2   & 
 8.5   & 
 14.0   &
 22.5   &
 17.8 \cr
$\Gamma^{bb}$  ($\mu$K/Jy)\tablenotemark{b} &
   250  & 230  & 240  & 240  & 240  & 270 \cr
\noalign{\smallskip\hrule\smallskip}
Rayleigh-Jeans: $S_{\nu}\propto \nu^{2}$  \cr
$\nu_e$  (GHz) &
 22.8  &  33.0  &  41.0  &  61.3  & 93.7  & 93.9  \cr
$\Delta\nu_e$ (GHz)  \tablenotemark{c}  &
  5.5 & 
  7.0  & 
  8.3  & 
  14.0 &
  23.0 &
  18.0 \cr
$\Omega^{bb}_m$ (deg$^2$)  \tablenotemark{d}   &
  0.77 & 
  0.44  & 
  0.26  & 
  0.12  &
  0.052  &
  0.044   \cr
$A^{bb}_e$ (m$^2$) \tablenotemark{e}  &
   0.72 & 
   0.60  & 
   0.67  & 
   0.66  & 
   0.61  &
   0.76   \cr
$g^{bb}$ (dBi) \tablenotemark{f}   &
 47.1  & 
 49.8   & 
 51.7   & 
 55.5   & 
 58.6   &
 59.3    \cr
\noalign{\smallskip\hrule\smallskip}
Dust: $S_{\nu}\propto \nu^{3.5}$   \cr
$\nu_e $ (GHz) &
  23.0 & 
  33.1  & 
  41.1  & 
  61.5  &
  94.2 &
  94.5   \cr
$\Delta\nu_e$ (GHz) &
   5.2 & 
   6.9  & 
   8.1  & 
   14.0  &
   19.0 &
   18.1   \cr
\noalign{\hrule\vskip1pt\hrule\smallskip}
}}}
\tablenotetext{a}{The full bandwidths associated with each waveguide
band. All beams pairs are left/right symmetric though there are
``upper'' (1 and 4) and ``lower'' (2 and 3) sets for W band with different
characteristics. All beam data are for the B
side. {\it These values are representative; the flight values will be different.}}
\tablenotetext{b}{$\Gamma$ is computed from equation~\ref{eq:jansk}.
The value is similar for all frequencies because the beam width is
proportional to the $\nu_e$ for the five bands.}
\tablenotetext{c}{The slightly lower effective frequencies reported in 
Jarosik {\it et al.} (2003) arise because the CMB is a thermal source and not a 
Rayleigh-Jeans emitter, as assumed here, and the CMB fills the beam and 
is not a point source.}
\tablenotetext{d}{The {\it measured} solid angle. It is close to but
not identical to $4\pi/g_m^{bb}$.} 
\tablenotetext{e}{The projected area of the primary is 1.54 m$^2$; the
physical area is 1.76 m$^2$. The effective area is derived from 
$A_{e}^{bb} =\lambda_{e}^2/\Omega_{m}^{bb}$.}
\tablenotetext{f}{The broadband gain from the models of the beam.}
\label{tab:measbeams}
\end{table*} 

\subsection{Window Functions}

The statistical characteristics of the CMB are most frequently expressed as an
angular spectrum of the form $l(l+1)C_l/2\pi$ \cite{Bond96} where $C_l$ is
the angular power spectrum of the temperature:
\begin{equation}
T(\theta,\phi) = \sum_{l,m}a_{lm}Y_{lm}(\theta,\phi),~~~C_l=<|a_{lm}|^2>.
\end{equation}
The spherical harmonic expansion is written so that $T(\theta,\phi)$
is real. The beam acts as a low pass filter on the angular
variations in $T(\theta,\phi)$, smoothing the sky over a characteristic
gaussian width of $\sigma_b=\theta_{fwhm}/\sqrt{8\ln(2)}$. The {\it
variance} of the time stream one would measure at the output of a
noiseless detector as the beam scans the sky is 
\begin{equation}
\approx\sum_{l}{(2l+1)C_l \over 4\pi } W_l
\end{equation}
where $W_l$ is the window function which encodes the beam smoothing. In
practice, one works with maps in which each pixel has been traversed in
multiple directions by an asymmetric beam. Then, one determines 
the variance of the ensemble of pixels as a function of $l$ and $\Delta
l$. The full window function for a separation of the beam centroids
of $\theta_{12}$ is
\begin{eqnarray}
W_{l}^{\gamma_1\gamma_2}(\theta_{12}) = \biggl( {1\over\Omega_A^2 }\biggr)
\int\int
P_l(\cos(\theta_{12}^\prime))
B_n(R_{\gamma_1}\vec\Omega_1,\vec\Omega_1^\prime)\times \nonumber\\
B_n(R_{\gamma_2}\vec\Omega_2,\vec\Omega_2^\prime)d\Omega_1^\prime
d\Omega_2^\prime\hspace{1truein}
\label{eq:winfcn}
\end{eqnarray}  
where $R_{\gamma}$ gives the orientation of the beams. The full
expression with orientable asymmetric beams has been used by
Cheng {\it et al.} (1994), Netterfield {\it et al.} (1997), Wu {\it et al.} (2000), 
and Souradeep \& Ratra (2001). For the zero lag window 
with symmetric beams, $W_l=B_l^2/\Omega_{A}^2$ where 
$B_l = 2\pi\int B_n(\theta)P_l(\cos\theta)d(\cos\theta)$
and $P_l$ are the Legendre polynomials.
If the beams are Gaussian, $W_l = \exp(-(l+1/2)^2\sigma_b^2) $. 

On the right side of Figure~\ref{fig:beam_maps} are shown the window 
functions for the B-side beams after they have been circularly symmetrized.
As the derived angular spectrum is directly multiplied by $W_l$
\cite{OSH}, some care must be taken in determining the window. {\it The
flight values will be different from those shown and will be quantified
with both models and in-flight beam maps.} 

\subsection{Practical Issues}

{\sl MAP} makes differential measurements and thus measures
the difference in power from radiation received from opposite
sides of the spacecraft. The relevant quantity
from which the maps are derived is
\begin{eqnarray}
\lefteqn{ W_D^{\gamma_1\gamma_2}(R_{\gamma_1}\vec\Omega_1-
R_{\gamma_2}\vec\Omega_2) = {1\over 2}  \int\int 
\bigg\lbrace   \alpha_A\eta_A A_{e,A}(\nu )f_A(\nu) } \hspace{3.25truein} 
\nonumber \\
\times S_{\nu}(\vec\Omega_1-\vec\Omega)\,
B_{n,A}(\nu ,R_{\gamma_1}\vec\Omega_1,\vec\Omega))-
\hspace{0.8in} \\
 \alpha_B\eta_B A_{e,B}(\nu )f_B(\nu)\,S_{\nu}(\vec\Omega_2-\vec\Omega)\,
B_{n,B}(\nu ,R_{\gamma_2}\vec\Omega_2,\vec\Omega)\,\bigg\rbrace
d\vec\Omega d\nu. \nonumber
\end{eqnarray}
for two different directions $\vec\Omega_1$ and $\vec\Omega_2$
for the 'A' and 'B' sides respectively.
Even if $\vec\Omega_1$ and $\vec\Omega_2$ are switched
(the A-side points to where the B-side was), $f_A = f_B$,  
$\alpha_A = \alpha_B$ 
(the phase switch is in the same position and the hybrids, waveguides,
and OMT are matched), and $A_{e,A} = A_{e,B}$,  $W_D$
is not zero. This is because $B_{n,A}$ and $B_{n,B}$ are not  
the same. The differential pairs are 
back-to-back and far from the ideal focus. 
In reality, $f_A\neq f_B$, $\alpha_A\neq \alpha_B$, 
and $A_{e,A} \neq A_{e,B}$, though to first order the small 
differences will be calibrated out. In the end, though, the different 
beams will have to be taken care of in the analysis. Because Jupiter is
so bright, each of the forty beam profiles will be independently mapped.
  
\section{Systematic Effects}

Systematic effects associated with the optics such as beam size, 
sidelobe levels, and the effective frequency directly affect the 
scientific interpretation of the data. In this section, systematic
effects associated with the stability and integrity of the optical
system are discussed.
 
The largest systematic effects are associated with infrared and 
microwave emission from the Sun, Earth, and Moon whose
properties, as viewed from L$_2$, are summarized in Table~\ref{tab:sem}.
 
\begin{table*}[t]
\caption{Parameters for Sun, Earth, and Moon, from L$_2$}
\small{ \vbox{
\tabskip 1em plus 2em minus .5em
\halign to \hsize {
     #\hfil & 
\hfil#\hfil &
\hfil#\hfil &
\hfil#\hfil \cr
\noalign{\hrule\vskip1pt\hrule\smallskip}
  	  &
  Sun  & 
  Earth  & 
  Moon   \cr
\noalign{\smallskip\hrule\smallskip}
 Temperature, T$_s$ (K)	  &
  6000                    &
  300                     & 
  250                     \cr 
 Distance from L$_2$, $d_s$ (m)	  &
  $1.5\times10^{11}$      &
  $1.5\times10^{9}$       & 
  $1.5\times10^{9}$       \cr 
 Radius, $r_s$ (m)	          &
  $7\times10^{8}$         &
  $6.4\times10^{6}$       & 
  $1.7\times10^{6}$       \cr
$I_{inc}^{ir}$ Flux at L$_2$ (W/m$^{2}$)\tablenotemark{a} &
  1600                    &
   $8.5\times10^{-3}$       &
   $2.8\times10^{-4}$       \cr
Solid Angle from L$_2$ (sr) &
   $6\times10^{-5}$                   &
   $6\times10^{-5}$       &
   $5.3\times10^{-6}$       \cr
$I_{inc}^{mw}/A$ in $\Delta\nu = 24~$GHz (W/m$^2$) \tablenotemark{b}&
 $2.4\times 10^{-8}$ &
 $1.0\times 10^{-9}$ &
 $1.1\times 10^{-10}$ \cr
$\rm T_{eff}$ (mK) \tablenotemark{c}&
 130  &
 5.5  &
 0.6 \cr
\noalign{\hrule\vskip1pt\hrule\smallskip}
}}}
\tablenotetext{a}{ $I = \sigma_B T^{4}(r_{s}/d_s)^2$ }
\tablenotetext{b}{For radiometric estimates, the flux in a microwave
frequency band is useful. It is given by
$I_{inc}= 2\pi \nu^2 k_B T_s ( r_s/d_s )^2\Delta \nu/c^2$. The values
are for 90 GHz.}
\tablenotetext{c}{The effective temperature is $\rm T_{eff} =
T_s(r_s/d_s)^2$.}
\label{tab:sem}
\end{table*}

\subsection{Thermal Variation of Optics}

The S/C is oriented so that the angle between the Sun and the
symmetry axis is constant during the spin and precession.
Thus, the thermal loading remains constant for long periods. 
The radiators are designed so that they are $2\fdg2$
degrees inside the Sun's shadow at all times. The reflectors too are
$>6\deg$ inside the Sun's shadow. Infrared emission
from the Earth and the Moon, however, can directly illuminate 
the primary as can be seen in Figure~\ref{fig:sideview}. 
In the following, conservative order-of-magnitude estimates are made of the
anticipated radiometric signal due to heating of the primary.

The primary is a complicated composite structure with multiple
thermal time constants and can only be treated roughly in
isolation. An input of thermal power, $\Gamma^t$, heats up the
mirror. The energy in turn is both immediately reradiated and flows to
other parts of the S/C where it is eventually reradiated. The energy
flow for both mechanisms is modeled as  
\begin{equation}
C_M {dT \over dt} + g^{th}(T-T_0) = \Gamma^t
\label{eq:therm}
\end{equation}
where the heat capacity is $C_M$ and $g^{th}$ is the thermal
conductance. The solution to equation~\ref{eq:therm} with a step function
change in incident power is:
\begin{equation}
T = T_0 + \Gamma^t(1-e^{-t/\tau})/g^{th}
\label{eq:trad}
\end{equation}
where $\tau = C_M/g^{th}$.  

The dominant conductive path is through the $0.0254$~cm 
thick composite skin of the reflector. The thermal conductivity of the
XN70 is $\kappa_s=0.19$~W cm$^{-1}$K$^{-1}$ at 70~K, more than 
twice that of stainless steel. The composite's density is
$\rho_s=1.8~{\rm gm\,cm^{-3}}$ and the specific heat capacity
is $c_s = 200~{\rm J\,kg^{-1}\,K^{-1}}$. The characteristic time 
for a heat pulse to propagate 
$l_h=20$~cm is $\tau_{cond}= l_h^2\rho_s c_s/\pi^2\kappa_s^2 \approx
80~$s, though the full solution is a series
expansion in time constants and depends on geometry \cite{Hildebrand76}.

The emitted radiation, for small temperature variations, is linearized so that 
$\delta I_{rad} = 4\epsilon_{ir}\sigma T_0^3\delta T $. The 
effective radiating area is 
the face of the primary with $A_{rad}=1.76~$m$^2$ for which the emissivity is
taken as $\epsilon_{ir}=0.1$ at $\lambda=30~\mu$m. (At
$\lambda=10~\mu$m and 300~K,  $\epsilon_{ir}=0.5$.) The conductance
is given by $g^{th}_{rad} = 4\epsilon_{ir}\sigma T_0^3A_{rad}=0.015~$W/K
at 70~K. The heat capacity is difficult to estimate
and depends on the coupling between 
the reflector surface and its backing structure. If the whole
5~kg reflector heats up, then $c_s\approx 250~{\rm J\,kg^{-1}\,K^{-1}}$, 
the heat capacity of
a typical composite, and $C_M\approx 1000~$J\,K$^{-1}$. If just 
the thin surface heats up,
then $C_M\approx 160~$J\,K$^{-1}$, leading to the range 
$3~{\rm h}<\tau_{rad}< 20~{\rm h}$.
Because $\tau_{cond} >> \tau_{rad}$, to a good approximation the
primary first thermalizes and then radiates.

To estimate $\Gamma^t = \epsilon_{ir} I_{inc}^{ir} A_{il}$ it is 
assumed that the Earth (Moon)
illuminates  $A_{il}\approx{\rm 0.15~m^2}$ ( $A_{il}\approx{\rm 1~m^2}$)
with an absorptivity of $\epsilon_{ir}\approx 0.1$. 
For the Earth, $\Gamma^t_E = 1.3\times
10^{-4}~$W and for the Moon, $\Gamma^t_M = 3\times 10^{-5}~$W.
The temperature increase for a long exposure
($>\tau_{rad}$) is $\Gamma^t/g^{th}_{rad}$. For the Earth, this is
$\Delta T_{phys}\approx 9~$mK and for the Moon it is 
$\Delta T_{phys}\approx 2~$mK.
The precise illumination depends on the orbit, spacecraft orientation, 
and blockage of the radiation by the
secondary as shown in Figure~\ref{fig:sideview} and so these estimates
should be considered conservative upper limits.

The rotation period of the satellite is 132~s. The reflectors have a
good view of the Earth and Moon over about $45\deg$ range, or for
about $\tau_{obs}=15~$s. In the limit that the heat is 
first conducted away from the heated area and later reradiated,
equation~\ref{eq:trad} gives a temperature change of
$\Delta T_{phys}=1.5~$mK for the Earth and $\Delta T_{phys}=0.35~$mK
for the Moon. The radiometric signal is proportional
to the thermal variation multiplied by the microwave emissivity
integrated over the beam response. In W-band, the optical response 
near the top of the primary, where the Earth illuminates it, 
is $\approx -5~$dB of the peak and so the expected radiometric
signal is $\approx 0.5~\mu$K. The Moon illuminates the central part 
of the primary and so the expected radiometric signal is $\approx
0.4~\mu$K. The primaries, secondaries, and radiators are 
instrumented with platinum resistive thermometers (PRTs) 
to detect a 0.5~mK change in
one rotation. These measurements and a detailed thermal model will be used
to model the on orbit performance.

The estimates above assume the same emissivity on both sides but 
variable thermal loading. If the microwave emissivity
of the two reflector systems is different and the temperatures of
both move up and down identically, then a signal will result.
If the temperature of the reflectors change together by 10~mK, 
which is conservative but not impossible, then the differential
signal changes by $1~\mu$K if $\delta\epsilon/\epsilon = 0.2$.

Finally, because the current distributions from each feed overlap on the 
primary, as shown in Figure~\ref{fig:curd}, temperature and emissivity
gradients will have a common-mode effect on all channels. This will aid
in characterizing any variations in the optics.  

\subsubsection{Emission From a ``Dirty'' Surface.}

The reflector surfaces are specified to be ``visibly clean.''
In contamination engineering, a visibly clean surface has $500<l_c
<700$, where $l_c$ is the ``cleanliness level.''
For example, a surface with one
$3~\mu$m by $30~\mu$m needle per square millimeter (0.01\%
obscuration) results in $l_c = 400$. If the particles are
spherical with the same fractional obscuration, $l_c =
250$. These surfaces would pass as visibly clean. On the other hand,
a surface covered with $20~\mu$m by $100~\mu$m needles with a 2\%
surface obscuration results in $l_c>1000$ and is clearly not
visibly clean. This corresponds to 1~g of graphite spread over the surface. 
Generally, the unaided eye can detect $50~\mu$m particles.  

During the integration and prelaunch phase, the optical 
surfaces are constantly purged
and kept clean. However, the fairing separation soon after launch can
produce a cloud of debris.
Even though the {\sl MAP} payload fairing was specially cleaned and inspected 
to minimize the possibility of contamination, it was estimated that there 
could be 1~g of contaminating material and one ${\rm 1\times 5~cm^2}$ piece of
fairing tape within the area of one reflector. The energetics and
geometry of the separation strongly disfavor much, if any, of this
material sticking to the primary. Of the possible contaminants 
shown in Table~\ref{tab:contams}, graphite is the most pernicious. 

Surface dust or debris increases
the microwave emissivity possibly resulting in a radiometric signal.
The magnitude of the signal depends on the material's coupling
to the surface. For a piece of ${\rm 1\times 5~cm}^2$ $0.05~$cm thick
Mylar tape, the most pessimistic case occurs when the contaminant
is attached at the center of the primary in a way such that its response
to temperature variations is instantaneous and its emissivity is unity 
in the infrared and
$\epsilon_{eff}=0.03$ at 90~GHz as shown in Table~\ref{tab:contams}.
The maximum radiometric temperature will be 
\begin{equation}
    \Delta T_{rad} = \epsilon_{eff}\Delta T_{object} 
{ A_{object}\over A_{il}}\approx 0.15~\mu{\rm K}.
\label{eq:emistemp}
\end{equation}
where $ A_{il}\sim 10^4~{\rm cm}^2$ is the illumination area of the beam, and 
$T_{object}\sim 10~$mK. 

Though there will be some graphite in the mix of debris, there is
far less than 1~gm. If, though, 1~gm of the 20~$\mu$m by 100~$\mu$m 
needles discussed above is distributed
over the surface there are roughly 10 particles per mm$^2$. Needles this 
close to the reflector will follow the surface temperature. Since they
are much smaller than a wavelength and in the node of the electric field 
their effective emissivity is small. The power in the standing wave as a
function of distance $z$ from the surface is proportional to
$\sin(4\pi z/\lambda)$. A $20~\mu$m diameter grain at 90 GHz
on the surface sees a reduction in power of $\approx 10^{-3}$ after
accounting for the radiation reflected to the scatterer from the mirror.
Thus we assume $\epsilon_{eff}\approx 0.001$.  The emission
temperature is then $\epsilon_{eff}T_{phys}A_{objects}/A_{il}\approx 0.02~\mu$K. 

We can never be certain how much contamination ends up on the surface.
The effective emission temperatures for large pieces of material and for
graphite grains are conservative upper limits. They are meant
to show {\sl MAP}'s immunity to contamination.
 
\subsection{Thermal Variation of the Radiators}

Sun light diffracts over the edge of the sun shield and
illuminates the radiator at a very low level. 
As the S/C spins this term is modulated, leading to a temperature
variation of the radiator and in turn the HEMT amplifiers.
Using GTD around the edge of the 
solar shield, a conservative estimate shows that no 
more than 0.1~W could be absorbed by the radiator.
The heat capacity of a radiator panel is $C_r=2500$ J\,K$^{-1}$ 
and so the temperature variation is
$\approx 0.1~{\rm W}C_r^{-1}\tau_{obs}=50~\mu$K for
$\tau_{obs}=1~$s. With an emissivity
of $<$1\%, the maximum radiometric signal is $1~\mu$K for perfect radiometric
coupling. Because the heat is injected 
at the outer edge of the radiator, far from the optics, and it is partially
reradiated, the resulting
radiometric signal is $<<1~\mu$K.  
 
\begin{table*}[t]
\caption{Possible contaminants at 90 GHz\tablenotemark{a}}
\small{ \vbox{
\tabskip 1em plus 2em minus .5em
\halign to \hsize {
     #\hfil & 
\hfil#\hfil &
\hfil#\hfil &
\hfil#\hfil &
\hfil#\hfil &
\hfil#\hfil \cr
\noalign{\hrule\vskip1pt\hrule\smallskip}
 Potential contaminant   &
   $\epsilon$    & 
   Re($\epsilon$ )   & 
   Im($\epsilon$ )   &
   $\alpha$ (cm$^{-1}$)   &
    Source\cr
\noalign{\smallskip\hrule\smallskip}
Graphite\tablenotemark{b}  &
$\epsilon = 700 + i(4.8\times 10^{-3}\nu^{1.13}  + 1.5\times 10^7/\nu)$ &
  700 &
  170000 &
  \dots &
 \cite{DraineLee84} \cr
Eccosorb CR110\tablenotemark{c,d}&
$\epsilon \approx 3.53 + i0.48\,\nu^{0.1}$ &
  3.53 &
  \dots &
   2.0 &
 \cite{Halpern86}\cr
Ice &
$\epsilon = 3.155 + i 2.8\times 10^{-4} \nu$&
  3.155 &
  0.026 &
  \dots &
 \cite{Koh97}\cr
Mylar &
$\epsilon \approx 2.82 +   i (0.15 + 1.7\times 10^{-4} \nu)$ &
  2.8 &
  0.17 &
  0.52 &
\cite{Page94}\cr
Silicates &
$\epsilon = 11.8 + i 1.1\times 10^{-3} \nu$&
  11.8 &
  0.1 &
  \dots &
\cite{DraineLee84} \cr
SiO$_2$ &
$\epsilon \approx 3.8 +  i 9\times 10^{-4} \nu^{1.51}$ &
  3.8 &
  0.8 &
  0.003 &
 \cite{MonSievers75}\cr
\noalign{\hrule\vskip1pt\hrule\smallskip}
}}}
\tablenotetext{a}{For all expressions in this Table, $\nu$ is in GHz.
The absorptions coefficient at 70~K, $\alpha$, is generally within 10\% 
of that at 290~K. The material
properties are quantified with the complex dielectric 
constant \cite{Jackson99,RW53}. }
\tablenotetext{b}{For a good though lossy or imperfect conductor, 
$\alpha = 1/\delta$ where $\delta = 1/\sqrt{\pi\mu\,\nu\sigma}$ is the 
skin depth. Thus $\alpha$ is expected to increase as $\nu^{1/2}$. The
dielectric constant is given by $\epsilon = 1 + i\sigma/(2\pi\,\nu)$.
Graphite has both a metallic and interband contribution.}
\tablenotetext{c}{Eccosorb CR110 is a castable resin microwave absorber
made by Emerson Cumming. NS43G/Hincom paint chips could have an 
absorptance as high as this.}
\tablenotetext{d}{For imperfect dielectrics, 
$\alpha = \pi \nu {\rm Im}(\epsilon )/c\,{\rm Re}(\epsilon )$
as long as the conduction currents are much less than the displacement current.
Sometimes the loss tangent is quoted,  
$\tan\delta = {\rm Im}(\epsilon)/{\rm Re}(\epsilon)$.}
\label{tab:contams}
\end{table*}

\subsection{Scattering of Light from the 
Sun, Earth, \& Moon by Contamination on Primary}

Microwave power from the Moon and Earth can scatter off
debris on the reflector and potentially produce a glint
as the S/C rotates. The form of the contamination is not known
and so several possibilities are considered.
The differential 
cross section \cite{Hulst,LL} for the scattering of unpolarized incident light
from an isotropic medium is
\begin{equation}
{d\sigma\over d\Omega } = \biggl({2\pi \nu \over c}\biggr)^4
|\alpha_p|^2 {(1+\cos^2\theta_s)\over 2},
\end{equation}
where $\alpha_p$ is the polarizability of the material, $V$ is the
volume, and $\theta_s$ is scattering angle. 
In the following, the angular term is taken as 3/4 and the
scattering from each grain is treated as isotropic. 

The polarizability of a needle and a sphere respectively are given by
\begin{equation}
\alpha_p^{needle} = { 1\over 4\pi }(\epsilon - 1)V ~~\&~~
\alpha_p^{sphere} = { 3\over 4\pi }
\biggr({\epsilon - 1\over \epsilon + 2}\biggl)V,
\label{eq:alphap}
\end{equation}
where $\epsilon$ is the dielectric constant and $V$ is the grain 
volume. The most efficient
scattering shape per unit volume is a needle.

If light with flux (power/area) strikes the scatterer,
the intensity at scattering angle $\theta_s$ and distance $r$ is 
\begin{equation}
I_{sca}(\theta_s) = {d\sigma\over d\Omega }I_{inc}^{mw}/r^2
\end{equation}
For a random set of incoherent scatterers with surface density $\sigma_N$,
the scattered power adds. Thus, the total scattered light is 
just $\sigma_N A$ times the above,
where $A$ is the area covered by the scatterers. 
This surface element subtends a
solid angle of $A \cos\theta_o/r^2$. Thus, the resulting surface brightness
is
\begin{equation}
B_{sca}(\theta_s) = \sigma_N {d\sigma\over d\Omega }I_{inc}^{mw}/
(\cos\theta_o\Delta \nu)
~~~~~{\rm W\, m^{-2}\,sr^{-1}\,Hz^{-1} }.
\end{equation}
The $\cos\theta_o$ in the denominator accounts for isotropic
scattering; the scattering is not Lambertian.  

The equivalent parabola concept (Table~\ref{tab:refl_params}) is used to compute
the measured signal. The scattered light from the top of the primary is
modeled as directly entering the feed. 
The Moon illuminates $\approx 0.2~$sr of the
primary at an angle from the feed of $\theta_o\approx20^{\circ}$ and
gain of 20 (13~dBi) as indicated in Figure~\ref{fig:sideview}.
The Earth, though hotter, illuminates less and is subdominant.
Because of the large uncertainty in the type of contaminant,
a full integration of the telescope response is not necessary.
The effective antenna temperature is given by:
\begin{eqnarray}
\label{eq:tscat}
T_{A} = {1\over 4\pi}\int {B_{sca}(\theta ) \lambda^2\over 2k_B}
g(\theta_o)d\Omega \hspace{1.0truein}  \\
=(3/16)\eta_{surf}\sigma_N \biggl({2\pi \nu\over c }\biggr)^4
|\alpha_p|^2 T_s(r_s/d_s)^2g(\theta_o )\Delta\Omega.\nonumber
\end{eqnarray}
where $r_s$, $d_s$, and $T_s$ are given in Table~\ref{tab:sem}.
The efficiency $\eta_{surf}$ encodes the fact that
the scatterers on a metal surface are at the node 
of the electric field. For lunar emission at 90 GHz, 
equation~41 becomes
$T_{A} = 7\times10^{9}\eta_{surf}\sigma_N |\alpha_p|^2$ in MKS units.

For the $20~\mu$m by $100~\mu$m needles at 90 GHz discussed above,
$\eta_{surf}\approx 10^{-3}$, $\sigma_N = {\rm 10^{7}m^{-2}}$,
 $|\alpha_p|^2/V^2\approx 2\times10^{9}$, and $V=3\times 10^{-14}~{\rm
m^3}$. The result is that $T_{A} = 120~\mu$K. If the mass of graphite
is held fixed but the shape of the grain is more spherical, 
$T_{A}$ is reduced to sub-$\mu$K
levels. Baring the pathological case of large needles, emission from
scattering off contaminants is negligible. 

\subsection{Nonuniform Coating of Primary}

It is possible that the primary may be coated with a thin layer of water or
hydrazine (fuel for the thrusters) which then freezes. Neither
of these absorbs microwave radiation but if they irregularly coat
the surface the variation in the thickness of material leads 
to a variation in the phase
of the wavefront in turn leading to a distortion of the beam. The phase delay
for a thickness $d$ is $\delta\phi=2dn/\lambda$. For example, if 1~g of water
freezes out over half the surface, $\delta\phi\approx 1.4\times 10^{-3}$.
This is negligible, more than two orders of magnitude smaller than 
the surface {\it rms} specification discussed in section~\ref{sec:surface} 

\subsection{Micrometeoroids}

\begin{figure*}
\epsscale{1.6}
\plotone{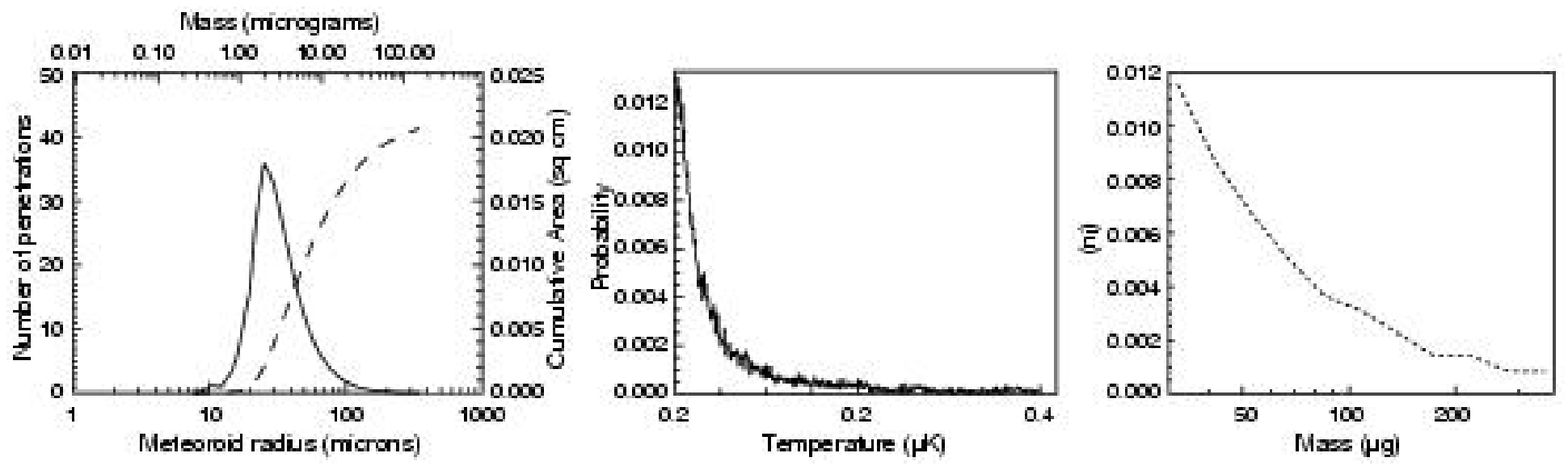}  
\caption{{\it Left:}The solid line shows the distribution of the number 
of penetrations per each of 50 meteoroid radius bins
for 15.8 m$^2$ area exposed (e.g., the Sun shield) for 27 months (the
nominal mission lifetime)\cite{jA98}. One assumes that each hit results
in a penetration. The dashed line is the cumulative area of the
penetrations. Note that most of the net
area comes from just a few particles. The radius as a function of 
mass is computed for a 
density of $2$~g/cm$^3$ and a spherical shape.{\it Center} The
normalized probability distribution for 35,000 realizations of $\Delta T$ at
the end of 27 months for two mirrors with $r_{hole}=a$. 
As a consequence of the differential measurement, the
mean value is zero. {\it Right:} The number of collisions per year for 
1.8~m$^2$ for each mass bin (dot). The probability of a hit by a 
particle of diameter larger 
than 200~$\mu$m hits is 0.2/m$^2$ per year, or 0.7 per primary per 27 months.}
\label{fig:meteoroid}
\end{figure*}

Micrometeoroids pelt the optics, thereby increasing their microwave
emissivity.
The particle distribution \cite{jA98,SteveBest} shown 
in Figure~\ref{fig:meteoroid} is
typical for Earth spacecraft, and believed to be representative of L$_2$.
The relative velocity to the S/C is $19$~km/sec and they may hit at any
angle. At the heavy end of the distribution, the particles have an energy of
$50$ J. Each penetration is modeled as a circular hole of diameter the
micrometeoroid multiplied by a factor. Holes in the sun shield
transmit a negligible amount of solar energy; holes in the reflectors
might act as blackbodies at 70 K, the temperature of the reflectors.

The micrometeoroids have a size distribution between 
15$~\mu$m and 700$~\mu$m and a mass between 0.004~$\mu$g and 400~$\mu$g. 
They are mostly olivines and silicates. The specific density for
particles under 1~$\mu$g is $2$~g/cm$^3$; for masses between 
1~$\mu$g and $0.01$~g, it is
$1$~g/cm$^3$; and it is $0.5$~g/cm$^3$ for more massive meteoroids.
In round figures, there is roughly $1$ hit/m$^2$ per 
year of a $100~\mu$m diameter particle and $100$ hits/m$^2$ 
of $10~\mu$m diameter particles. 

To compute the flux that scrubs the reflectors, the values in 
Figure~\ref{fig:meteoroid} are scaled to one month and a surface of
$1.8$~m$^2$,  corresponding to one primary reflector. The holes (places where
the Al is removed) are modeled as
circular apertures that transmit 90 GHz. For a 
300 $\mu$m radius particle, $ka=0.57$ ($a$ is the radius) thus some care
must be taken in computing the transmission coefficient. Landau's
expression \cite{Jackson99} is used even though it is imprecise 
(though sufficient) near
$ka\sim1$. The power transmission coefficient is  
\begin{equation} 
\label{eq:landau}
\tau_{trans} (ka) = 1 - \frac{1}{2 k a} \int_0^{2 k a} J_0 (t) \, dt
\approx \frac{{(k a)}^2}{3} \quad \mbox{for} \quad ka \ll 1
\end{equation}

A perfect mirror, with holes of radii $a_i$, $i\in\{1,n\}$, has an
effective temperature given by: 
\begin{equation}
T_{A} = \frac{ \left( {\epsilon}_{Al} \left( A_{refl} - \sum_i
\pi a_i^2 \right) + \sum_i \tau_{trans} (a_i) \pi a_i^2 \right)} 
{A_{refl}} T_{refl}
\end{equation}
where $\tau_{trans}$ is the transmission coefficient of a hole, $\epsilon$
is the emissivity of the coated surface, $T_{refl}$ is the reflector 
temperature and $A_{refl}$ is its area. 
We conservatively model the holes as perfect, though small, emitters.
This model is valid in
the limit of $\sum_i\pi a_i^2<<A_{refl}$ 
and $\tau_{trans}>\epsilon_{Al}$. For each successive hit,
\begin{eqnarray}
\Delta T_{A} = \frac{\left(- {\epsilon}_{Al}+
\tau_{trans}(a_i) \right) \pi a_i^2} {A_{refl}}\, T_{refl} 
\end{eqnarray}

At 94~GHz, $\tau_{trans} \approx \epsilon_{Al}$ for particles with 
$a<20~\mu$m and the simple model breaks down. However,
these particles contribute negligibly to the overall emissivity.
The primaries and the particle flux on them are assumed to be independent.
For each mass bin of index $i$ and each month, the number $n_i$ of
collisions is randomly chosen for each reflector according to the
Poisson distribution. The change in $T_{A}$ for one reflector
is given by $n_i \times \Delta T_{A} (a_i)$, where $a_i$ is the radius of the
particles belonging to the particular mass bin. After adding the
individual contributions of all of the bins and taking the difference
between reflectors, the result is recorded and the process starts again 
with a new month. 

To assess the net differential signal produced by the collisions, one
simulates many {\sl MAP} missions as shown in Figure~\ref{fig:meteoroid}.
The standard deviation of the data is $0.25~\mu$K. From the 
cumulative probability distribution one finds that with 95\%
probability, the difference in temperatures will be smaller than 
$0.5~\mu$K. Most evidence suggests the holes will be
bigger than the particle size. To bound the problem, consider
$r_{hole}=5a$. As $\Delta T_{A}\propto\sim~a^4$ in the $ka<<1$ limit, one expects
a distribution 625 times wider. However, for the largest particles,
$ka\approx 3$,  and the small size approximation
in equation~\ref{eq:landau} is not valid. The net effect is that the
95\% upper limit on an offset is $140~\mu$K.

This offset is virtually undetectable. 
If {\sl MAP} were not differential,
the net signal from micrometeoroids would not be much different. This
is because the damage to the reflector is dominated by relatively 
few encounters with 100 $\mu$m size particles. With two large reflectors, the
cross section doubles and is not compensated by the differential measurement.

In the right hand side of Figure~\ref{fig:meteoroid} is plotted
the average number of collisions per month for the heaviest mass
bins. These collisions produce large holes but are rare. Most
important, the contribution to the
total hole area from the largest particles is not increasing as one
moves to yet larger particles, indicating that if the distribution is
accurate, $140~\mu$K is a reasonable bound on the change in offset.
This effect would most likely be masked by other radiometer effects and
would, in any case, be strongly suppressed by the mapmaking algorithm
and scan strategy.

\section{Conclusions}
We have designed a differential optical system for performing precise and 
accurate measurements of the anisotropy in the cosmic microwave background
with an angular resolution of $< 0\fdg23$. All major components of the 
system have been measured and modeled in detail. Such a 
characterization is required in order to give us confidence in
the scientific conclusions we derive from {\sl MAP}. In addition,
we have presented estimates of a number of systematic effects and have shown
that in all cases their radiometric contribution to the celestial
signal is negligible. 
    
\section{Acknowledgements}
The development of the {\sl MAP} optics started in 1993 and has involved 
many people in addition to the authors. At Princeton, N. Butler, 
W. Jones, and S. Bradley
wrote senior theses on various aspects of the development; C. Bontas
worked on the calculations for the micrometeroid and surface
deformations; A. Marino, D. Wesley, C. Steinhardt, C. McLeavey, C. Dumont, 
M. Desai, M. Kesden, A. Furman, O. Motrunich, R. Dorwart, E. Guerra,
\& C. Coldwell worked on modeling and testing various components;
C. Sule, and G. Atkinson worked on building optical components;
and S. Dawson, A. Qualls, and K. Warren kept the Princeton effort
running smoothly.
At UBC, M. Jackson measured the fine scale surface deformations and
C. Padwick measured the cryogenic properties of thin aluminum coatings. 
J. Heaney of Swales along with S. Dummer and R. Garriott 
of SOC helped define and 
worked especially hard to produce a surface that met the specifications. 
D. Neverman at PCI led the team that built the TRS. 
S. Best and J. Anderson provided the
micrometeoroid test data and initial meteoroid flux calculations, 
respectively. The NASA/GSFC team, led by L.
Citrin, the project manager, worked long hours to make the {\sl MAP}
optics a reality. C. Trout-Marx independently verified the design with Code V;
S. Seufert and K. Hersey ran the beam mapping effort and K. Hersey verified
and ran the beam prediction code at GSFC; 
T. VanSant led the materials analysis team, supported 
by a large group including B. Munoz, C. He, L. Wang, and C. Powers;
the reflector and blanket surface composition characterization was 
additionally supported by L. Bartusek, L. Kauder, R. Gorman, and W. Peters;
S. Glazer led the thermal predictions group, with key support from 
D. Neuberger; P. Mule led the STOP 
verification effort (building on the work of J. McGuire);
L. Lloyd and P. Trahan did the PRT harness fabrication and 
installation; A. Herera, H. Sampler, D. Osgood, C. Aviado, M. Hill, 
D. Schuster, T. Adams, and M. Holliday made sure that the
optics were aligned and blanketed correctly;
W. Chen, S. Ngo, and J. Stewart
led the mechanical design team at GSFC, with key 
support from B. Rodini, J. Parker, M. Schoolman, and T. Driscoll;
E. Packard led the facility support activities;
M. Jones ensured quality assurance;
and A. Crane, with support from A. Herrera, N. Dahya, R. Hackley,
and M. Lenz, led the TRS I \& T effort.
The GTD calculations, especially equation~\ref{eq:gtd2},
were done with J. Mather when he was part of the team 
(though any errors here are independent of his work).
Comments at early design reviews by D. Fixsen, P. Timbie, 
and P. Napier and the
input from a review led by J. Mangus were especially helpful.
We also thank B. Griswold of NASA/GSFC for his work on the figures.
The modeling of the optics was done with assistance
and code from YRS Associates: Y. Rahmat-Samii, W. Imbriale, and
V. Galindo. Finally, we thank an anonymous reviewer whose comments 
improved the paper.
This research was supported by the {\sl MAP} Project
under the NASA Office of Space Science and Princeton University. 
More information about {\sl MAP} may be found at http://map.gsfc.nasa.gov.

\end{document}